\documentclass[a4paper,11pt]{article}
\pdfoutput=1
\usepackage{jcappub}
\usepackage{mathtools}
\usepackage[T1]{fontenc}
\usepackage{amsmath}
\usepackage{graphicx}

\graphicspath{{plots/}}

\newcommand{\gs}{g_\star}
\newcommand{\gss}{g_{\star s}}
\newcommand{\Trh}{T_\text{rh}}
\newcommand{\Tmax}{T_\text{max}}
\newcommand{\Hrh}{H_\text{rh}}
\newcommand{\arh}{a_\text{rh}}
\newcommand{\Mpl}{M_\text{Pl}}
\newcommand{\sv}{\langle \sigma v\rangle}

\def\beq{\begin{equation}\begin{aligned}}
\def\eeq{\end{aligned}\end{equation}}

\title{Boltzmann Suppressed\\ Ultraviolet Freeze-in}
\author[a]{Nicolás Bernal,}
\author[b]{Sagnik Mukherjee,}
\author[b]{and James Unwin}

\affiliation[a]{New York University Abu Dhabi,\\
PO Box 129188, Saadiyat Island, Abu Dhabi, United Arab Emirates.}
\affiliation[b]{Department of Physics,  University of Illinois Chicago, Chicago, IL 60607, USA.}

\emailAdd{nicolas.bernal@nyu.edu}
\emailAdd{smukhe34@uic.edu}
\emailAdd{unwin@uic.edu}

\abstract{If the dark matter mass $m$ exceeds the maximum temperature of the Universe ($\Tmax< m$) then its production rate will be Boltzmann suppressed. The important implications of this Boltzmann suppression have been explored for dark matter freeze-in via renormalizable operators. Here we extend these considerations to the case of ultraviolet (UV) freeze-in for which freeze-in proceeds via non-renormalizable operators. The UV freeze-in variant has a number of appealing features, not least that a given effective field theory can describe a multitude of UV completions, and thus such analyses are model agnostic for a given high dimension freeze-in operator. We undertake model independent analyses of UV freeze-in for portal operators of general mass dimensions. Subsequently, we explore a number of specific examples, namely, Higgs portals, bino dark matter, and gravitino dark matter. Finally, we discuss how significant differences arise if one departs from the standard assumptions regarding inflationary reheating (i.e. transitions from an early matter dominated era to radiation domination). As a motivated example we examine the implications of early kination domination. Boltzmann suppressed UV freeze-in is well motivated and permits a number of compelling scenarios. In particular, we highlight that for $\Tmax \sim 1$~TeV it is feasible that the freeze-in mechanism is entirely realized within a couple of orders of magnitude of the TeV scale, making it experimentally accessible in contrast to traditional freeze-in scenarios.}
\begin{document}
\begin{flushright}
\end{flushright}
\maketitle

%%%%%%%%%%%%%%%%%%%%%%%%%%%%%%%%%%%%%
\section{Introduction}
%%%%%%%%%%%%%%%%%%%%%%%%%%%%%%%%%%%%%
The dark matter relic abundance could have been set in the first few moments after inflationary reheating~\cite{Dolgov:1989us, Traschen:1990sw, Kofman:1994rk, Kofman:1997yn, Barman:2025lvk}. An early exploration of this idea was outlined in the seminal paper of Giudice, Kolb, and Riotto~\cite{Giudice:2000ex}, which highlighted, among other insights, that dark matter production can proceed even if the mass of dark matter exceeds the reheating temperature of the Universe $\Trh$. In particular, such a scenario would typically lead to a Boltzmann suppression in the production rate. This possibility has recently been revisited in the ``Freeze-in at Stronger Coupling'' scenario of Cosme, Costa, and Lebedev~\cite{Cosme:2023xpa}, which has generated a variety of related studies by various authors~\cite{Silva-Malpartida:2023yks, Gan:2023jbs, Cosme:2024ndc, Koivunen:2024vhr, Arcadi:2024wwg, Boddy:2024vgt, Silva-Malpartida:2024emu, Arcadi:2024obp, Lebedev:2024mbj, Bernal:2024ndy, Lee:2024wes, Belanger:2024yoj, Arias:2025nub, Lebedev:2025dhq, Khan:2025keb, Bernal:2025osg, Chanda:2025bpl, Bernal:2025qkj, Arias:2025tvd}. This is partly motivated by the improved prospects for observational signals, compared to typical models of dark matter freeze-in~\cite{McDonald:2001vt, Choi:2005vq, Kusenko:2006rh, Petraki:2007gq, Hall:2009bx, Bernal:2017kxu}. Detection prospects are maximized for lower reheating temperatures (that is, $\Trh\sim$ MeV-GeV).

Previous works that concern ``Freeze-in at Stronger Coupling'' have been focused on specific models of dark matter, in particular the Higgs portal~\cite{Silva-Malpartida:2023yks, Silva-Malpartida:2024emu, Khan:2025keb, Arcadi:2024wwg, Lebedev:2024mbj, Bernal:2025osg, Bernal:2025qkj}, light dark photons~\cite{Bernal:2024ndy, Boddy:2024vgt, Arias:2025tvd}, $Z'$ mediators~\cite{Arcadi:2024obp, Belanger:2024yoj} or sterile neutrinos~\cite{Koivunen:2024vhr}. Here we reexamine the prospect of Boltzmann-suppressed freeze-in but rather within the model-independent framework of Ultraviolet Freeze-in~\cite{Elahi:2014fsa}. In UV freeze-in the dark matter and Standard Model thermal bath only communicate via non-renormalizable operators, which leads to several attractive features:
\begin{itemize}
    \item For relatively low reheat temperature, one typically expects any mediator states can be integrated out, naturally yielding an effective field theory (EFT) description.
    \item Generically, non-renormalizable operators will connect distinct sectors of the low-energy theory. Thus, UV freeze-in is a generic production mechanism.
    \item UV freeze-in is model agnostic, since a given EFT will describe myriad UV completions.
\end{itemize}

For freeze-in models that involve mediators,\footnote{For dark matter and visible sector coupled directly in a renormalizable manner that does not involve a mediator, then freeze-in production is IR dominated at temperatures $T\sim m$. The quintessential example being the mixed quartic $|H|^2|\phi|^2$ between the Higgs $H$ and scalar dark matter $\phi$, provided $\Tmax > m_h \simeq 125$~GeV. This Higgs portal was studied for Boltzmann-suppressed freeze-in in Refs.~\cite{Cosme:2023xpa, Silva-Malpartida:2023yks}.} production can be either IR-dominated or UV-dominated; in the latter case, freeze-in predominantly occurs at the highest temperatures of the thermal bath, i.e.~around the reheat temperature $T\sim \Trh$.  The determiner between IR freeze-in and UV freeze-in is whether the mediator can be produced on-shell, which depends on whether the thermal bath temperature ever exceeds the mediator mass $M$. For freeze-in via a mediator $M \ll \Tmax$ implies IR freeze-in, while $M >\Tmax$ leads to UV freeze-in.

In the case where inflationary reheating is well approximated as instantaneous, then $\Tmax \simeq \Trh \simeq \sqrt{\Gamma\, \Mpl}$ where $\Gamma$ is the decay rate of the inflaton. In particular, if the inflaton does not decay instantaneously, then while inflationary reheating will still terminate with the visible bath reaching a temperature of $\Trh$, before this point the thermal bath may be heated above $\Trh$ to a temperature $\Tmax$ (which we define later). Any physical processes that occur at temperatures above $\Trh$ will be impacted by the reheating process as the inflaton energy is transferred to the thermal bath~\cite{Giudice:2000ex} which can lead to significant deviations. For implications for freeze-in beyond the instantaneous decay approximation, see Refs.~\cite{Giudice:2000ex, Garcia:2017tuj, Bernal:2019mhf, Bernal:2025fdr}.

As highlighted above, the same EFT of UV freeze-in can describe a multitude of IR freeze-in theories, with all relevant IR quantities (masses and couplings) combined into an EFT cut-off scale $\Lambda$. Notably, for processes that occur at energies above $\Lambda$, the EFT will breakdown, and one is forced to specify a UV completion. The simplest example of freeze-in via a mediator is perhaps the Higgs portal to fermion dark matter $\psi$. The EFT description of this fermionic Higgs portal is via the dimension five operator $|H|^2\bar{\psi}\psi$. This can be UV completed in numerous ways, the most typical manner is to introduce a new scalar mediator $\varphi$ that couples to both the Higgs doublet $\varphi |H|^2$ and the dark matter $\varphi \bar{\psi}\psi$, or similar.

Previous papers on Boltzmann-suppressed freeze-in via mediators~\cite{Silva-Malpartida:2023yks, Gan:2023jbs, Cosme:2024ndc, Koivunen:2024vhr, Arcadi:2024wwg, Boddy:2024vgt, Silva-Malpartida:2024emu, Arcadi:2024obp, Lebedev:2024mbj, Bernal:2024ndy, Lee:2024wes, Belanger:2024yoj, Arias:2025nub, Lebedev:2025dhq, Khan:2025keb, Bernal:2025osg, Chanda:2025bpl, Bernal:2025qkj, Arias:2025tvd} have largely focused on the case that $M < \Tmax$, thus analyzing some specific class of renormalizable models. Here we study the converse scenario and examine the case with both $m$ and $M > \Tmax$, which allows us to frame our discussions in the EFT language. As such, we will require that the EFT cutoff  satisfies $\Lambda>\Tmax$. We will show how the EFTs we study here match to specific UV completions appearing in articles in the literature. To reiterate our main point, the application of an EFT description both simplifies the analyzes and allows one to make more model-independent statements regarding Boltzmann suppressed dark matter freeze-in. 

We also highlight here that freeze-in can be (final state) p-wave suppressed once the dark matter mass exceeds the temperature.  This is the first time p-wave suppression in freeze-in has been studied, although we find it has only a mild impact on the final results.

This paper is structured as follows: In Section~\ref{sec2} we derive the freeze-in yield in the Boltzmann suppressed regime of UV freeze-in assuming standard inflationary cosmology (both with and without the instantaneous reheating assumption). In Section~\ref{sec3} we proceed to outline a number of models which naturally realize this scenario, focusing on the classic Higgs portals, and well-motivated supersymmetric scenarios in which the lightest supersymmetric particle is the dark matter. Specifically, we examine implications for bino and gravitino dark matter. In Section~\ref{sec4} we discuss generalizations to non-standard cosmology, assuming general scaling for both Hubble and the temperature of the thermal bath. We then specialize to early kination domination and highlight the differences to the conventional assumption of matter domination prior to inflaton decays. We present our concluding remarks in Section~\ref{sec5}.

%%%%%%%%%%%%%%%%%%%%%%%%%%%%%%%%%%%
\section{UV Freeze-in at Stronger Coupling} \label{sec2}
%%%%%%%%%%%%%%%%%%%%%%%%%%%%%%%%%%%
In this section we outline the parametric scaling of the dark matter relic abundance assuming it is produced via UV freeze-in in the Boltzmann suppressed regime. This derivation is performed in a model-independent manner with respect to the freeze-in operator.

%%%%%%%%%%%%%%%%%%%%%%%%%%%%%%%%%%%%%%
\subsection{UV freeze-in}
%%%%%%%%%%%%%%%%%%%%%%%%%%%%%%%%%%%%%%
The vanilla picture of early Universe cosmology holds that, at the end of inflation, the inflaton energy density dominates the Universe and leads to a period of matter domination. Subsequently, perturbative decays of the inflaton to Standard Model degrees of freedom\footnote{We will assume that the inflaton does not appreciably couple to dark matter, since this leads to an initial non-zero abundance of dark matter and also implies an inflaton mediated freeze-in operator. See Refs.~\cite{Adshead:2016xxj, Hardy:2017wkr, Cosme:2024ndc} for a relevant discussion.} transition the Universe into a radiation-dominated epoch. With this assumption of standard cosmology, the expansion rate of the Universe before ($a<\arh$) and after ($a>\arh$) inflationary reheating (at the scale factor $a=\arh)$ is given by~\cite{Allahverdi:2020bys}
\beq
    H(a) \simeq \Hrh \times
    \begin{dcases}
        \left(\frac{\arh}{a}\right)^{3/2} &\text{ for } a \leq \arh\,,\\[8pt]
        \left(\frac{\arh}{a}\right)^2 &\text{ for } \arh \leq a\,,
    \end{dcases}
\eeq
and the temperature of the Standard Model thermal bath evolves as
\beq
    T(a) \simeq \Trh \times
    \begin{dcases}
        \left(\frac{\arh}{a}\right)^{3/8} &\text{ for } a \leq \arh\,,\\[8pt]
        \left(\frac{\arh}{a}\right) &\text{ for } \arh \leq a\,.
    \end{dcases}
\eeq
At the end of the reheating era, when the Universe enters into radiation domination, the Hubble rate is given by
\begin{equation}
    \Hrh \equiv H(\Trh) = \frac{\pi}{3}\, \sqrt{\frac{\gs(\Trh)}{10}}\, \frac{\Trh^2}{\Mpl}\,,
\end{equation}
where $\Mpl \simeq 2.4 \times 10^{18}$~GeV is the reduced Planck mass, and $\gs(T)$ corresponds to the number of relativistic degrees of freedom contributing to the Standard Model energy density. 

In freeze-in scenarios, it is typically assumed a negligible initial abundance of dark matter: $n_{\rm DM}(\Trh)\simeq0$.\footnote{The initial number density $n_{\rm DM}(\Trh)$ does not need to be zero, but it is assumed to be subdominant with respect to the abundance produced by the FIMP mechanism. Generically inflationary reheating or gravitational production \cite{Kolb:2023ydq} will result in a non-zero $n_{\rm DM}(\Trh)$. We assume here that $n_{\rm DM}(\Trh)$ can be made appropriately negligible. See e.g. Refs.~\cite{Adshead:2016xxj, Hardy:2017wkr, Kolb:2023ydq, Cosme:2024ndc} for relevant discussions.} Dark matter is subsequently produced by interactions in the Standard Model thermal bath. The temperature evolution of the dark matter number density $n_{\rm DM}(T)$ due to freeze-in can be described by the Boltzmann equation 
\beq \label{eq:BE0}
    \frac{{\rm d}n_{\rm DM}}{{\rm d}t} + 3\, H\, n_{\rm DM} \simeq \sv n_\text{eq}^2 
\eeq
where $n_\text{eq}(T) \sim T^3/\pi^2$ is the Standard Model bath equilibrium number density (we neglect the factor of bath degrees of freedom, since these cancel out in the final result).

In the above, $\sv$ denotes the dark matter production cross section. In the cases where the dark matter production from the Standard Model thermal bath can be described by $2\to2$ processes via an effective operator, the dark matter thermally-averaged production cross section is given by (see Appendix~\ref{sec:app} for derivations) 
\beq \label{sv}
    \sv \sim
    \begin{dcases}
        \frac{T^n}{\Lambda^{n+2}} &\text{ for } T \gg m\,,\\[6pt]
        \frac{m^n}{\Lambda^{n+2}} \left(\frac{m}{T}\right)^{2-L}\, e^{-\frac{2\, m}{T}} & \text{ for } m \gg T\,,
    \end{dcases}
\eeq
where $m$ is the dark matter mass, $\Lambda$ is the scale of new physics that induces the effective operator. The temperature dependence depends on $n$, which corresponds to an operator of mass dimension $5+\frac{n}{2}$ for $n\in\mathbb{Z}^{\rm even}$. Observe, in particular, the (namesake) Boltzmann suppression in the production cross section for $m\gg T$.

In addition, for $m\gg T$ the form of the cross section also depends on the lowest orbital angular momentum allowed for the final state pair $L$. The $L$-dependent suppression is analogous to $v_{\rm rel}^2$-suppression in freeze-out. Recall that $p$-wave suppression in freeze-out is evident from a partial wave expansion on the incoming states. For freeze-in with $T\ll m$ a partial wave expansion on the final states can yield an analogous $p$-wave suppression. In both cases, the Lorentz structure of the operator determines whether such a suppression occurs. For most freeze-in scenarios, the leading allowed process will be either $L=0$ ($s$-wave) or $L=1$ ($p$-wave). Given exponential suppression, the difference between $L=0$ and $L=1$ tends not to be very significant. Notably, while the production cross-sections in the two cases can differ by a factor of $\sim 10$, this corresponds to small $\mathcal{O}(1)$ deviations in our plots. Since the differences are not apparent in the (log-scale) plots, we consider only the $L=0$ case in this section and discuss how the $L=1$ results (mildly) vary in Appendix~\ref{sec:app}.

To match the entire observed dark matter relic density, it is required that
\begin{equation} \label{eq:observed_DM}
    m\, Y_0 = \frac{\Omega h^2\, \rho_c}{s_0\, h^2} \simeq 4.3 \times 10^{-10}~\text{GeV},
\end{equation}
where $Y_0$ is the asymptotic value of the dark matter yield $Y(T) \equiv n(T)/s(T)$ at low temperatures, defined with respect to the Standard Model entropy density
\begin{equation}
    s(T) = \frac{2\pi^2}{45}\, \gss\, T^3,
\end{equation}
with $\gss(T)$ corresponding to the number of relativistic degrees of freedom that contribute to the Standard Model entropy. In addition, $s_0\equiv s(T_0) \simeq 2.69 \times 10^3$~cm$^{-3}$ is the present entropy density~\cite{ParticleDataGroup:2024cfk}, $\rho_c \simeq 1.05 \times 10^{-5}~h^2$~GeV/cm$^3$ is the critical energy density of the Universe and $\Omega h^2 \simeq 0.12$ is the observed dark matter relic abundance~\cite{Planck:2018vyg}.

%%%%%%%%%%%%%%%%%%%%%%%%%%%%%%%%%%%%%
\subsection{Instantaneous reheating}
%%%%%%%%%%%%%%%%%%%%%%%%%%%%%%%%%%%%%
The details of the cosmological history can significantly affect the dynamics and parametrics of dark matter freeze-in. In particular, UV freeze-in is sensitive to the details of inflationary reheating. Here we shall assume the conventional picture of inflationary reheating in which the coherently oscillating inflaton field dominates the energy density of the Universe leading to a matter dominated era, before subsequently decaying. We shall first consider the case in which the entire inflaton energy density is transferred to radiation instantaneously at $H\simeq\Gamma$. Then in the next subsection, we consider the case in which this process cannot be modeled as instantaneous and discuss the implications. We return to discuss non-standard cosmological scenarios in Section~\ref{sec4}, for instance, deviating away from the assumption that the Universe is matter dominated prior to inflationary reheating.

%%%%%%%%%%%%%%%%%%%%%%%%%%%%%%%%%%%%%%%%%%%%%%%%%%%
\begin{figure}[t!]
    \def\sepf{0.328}
    \centering
    \includegraphics[width=\sepf\columnwidth]{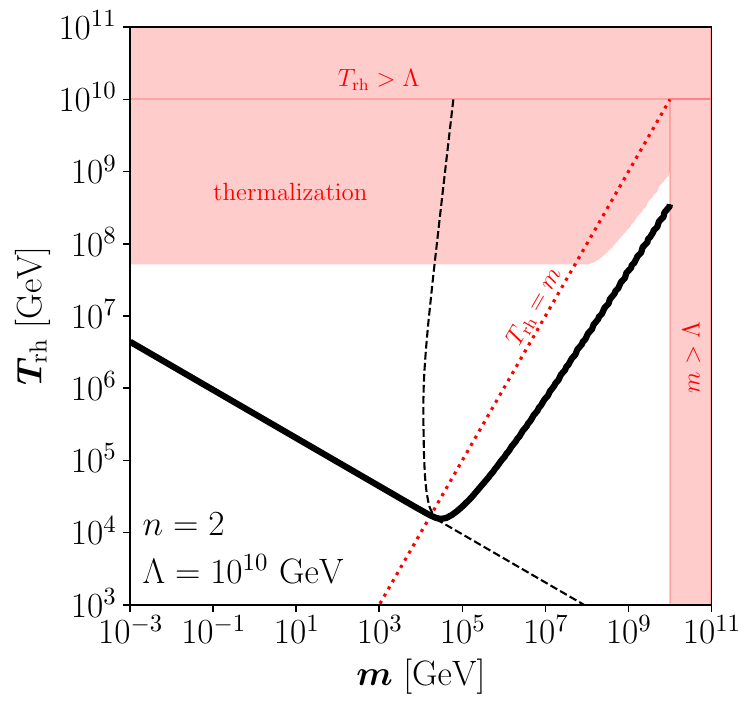}
    \includegraphics[width=\sepf\columnwidth]{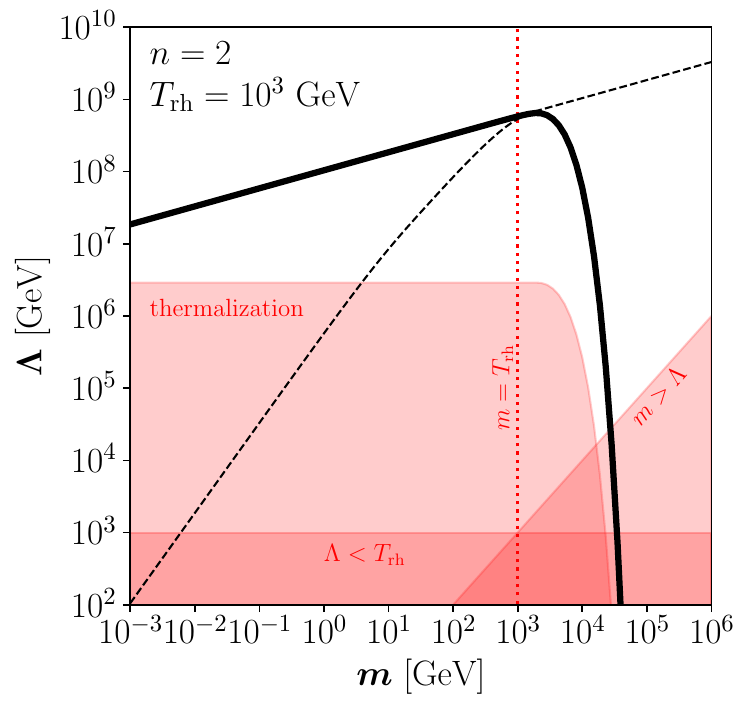}
    \includegraphics[width=\sepf\columnwidth]{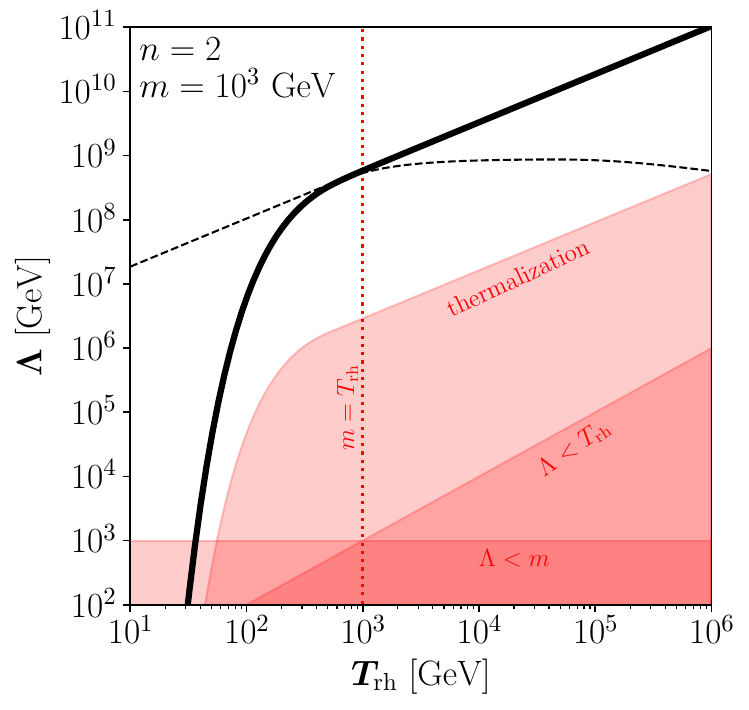}
    \caption{{\bf Instantaneous reheating.} The thick black solid lines show the parameter values that reproduce the observed dark matter abundance (that is Eq.~\eqref{eq:Yafter}). The thin black dashed correspond to the relativistic and non-relativistic approximations (cf. Eq.~\eqref{eq:Yafter-app}), for $n=2$ (mass dimension six freeze-in operator). The transition between the two regimes is depicted by the dotted red lines $m = \Trh$. The red bands correspond to the regions where the EFT approach breaks ($m > \Lambda$ and $\Trh > \Lambda$), and the thermalization limit (i.e.~in this region the dark matter enters thermal equilibrium with the Standard Model bath).
    \label{fig:plots}}
\end{figure}
%%%%%%%%%%%%%%%%%%%%%%%%%%%%%%%%%%%%%%%%%% 

When the instantaneous reheating approximation holds, the Standard Model entropy is always conserved and $\Trh$ is the highest temperature reached by the Standard Model thermal bath. In that case, we can calculate the late-time dark matter yield by integrating Eq.~\eqref{eq:BE0} from the point of reheating at scale factor $\arh$ to today at $a_0\equiv a(T_0)$ as follows
\beq \label{eq:Yafter}
    Y_0 &= \frac{45}{2\pi^2\, \gss(T_0)}\, \frac{1}{a_0^3\, T_0^3} \int_{\arh}^{a_0} \frac{a^2}{H(a)}\, n_\text{eq}^2\, \sv\, {\rm d}a\,. 
\eeq
We take the model-independent production cross section of Eq.~\eqref{sv} for $L=0$ (see Appendix~\ref{sec:app} for $L>0$) and perform the integration to obtain the piecewise-defined dark matter yield 
\beq \label{eq:Yafter-app}
    Y_0  &\simeq \frac{135}{2\pi^7\,\gss}\, \sqrt{\frac{10}{\gs}}\, \frac{1}{1+n}\, \frac{\Mpl\, \Trh^{1+n}}{\Lambda^{2+n}} \times
    \begin{dcases}
        1 & \text{for } m \ll \Trh\,,\\[3pt]
        \frac{1+n}{2} \left(\frac{m}{\Trh}\right)^{1+n} e^{- \frac{2 m}{\Trh}} & \text{for } \Trh \ll m\,,
    \end{dcases}
\eeq
For $m \ll \Trh$, the standard result is recovered~\cite{Elahi:2014fsa}. For $m \gg \Trh$ dark matter production is exponentially suppressed by two Boltzmann factors $e^{-m/\Trh}$ if $L=0$, or by stronger suppression if $L>0$.

%%%%%%%%%%%%%%%%%%%%%%%%%%%%%%%%%%%
  \begin{figure}[t!]
    \def\sepf{0.328}
    \centering
    \includegraphics[width=\sepf\columnwidth]{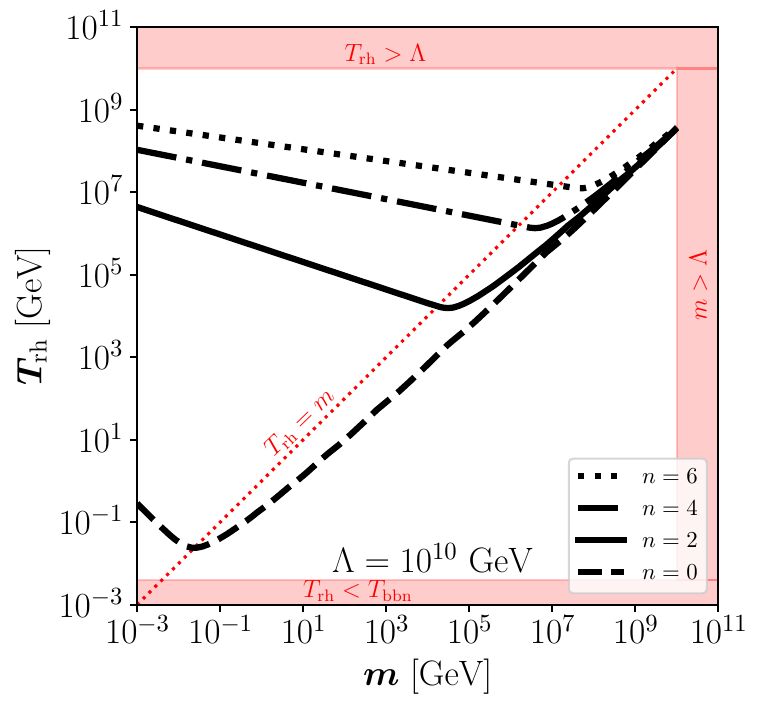}
    \includegraphics[width=\sepf\columnwidth]{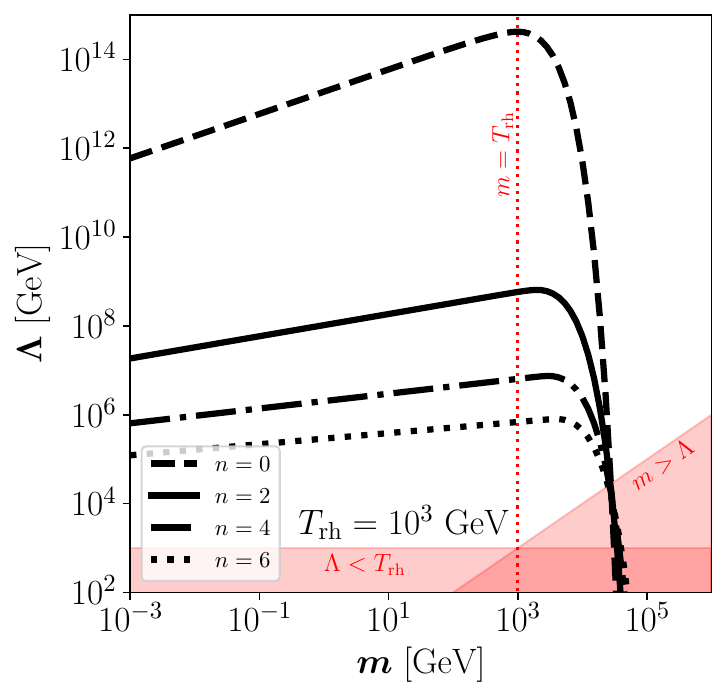}
    \includegraphics[width=\sepf\columnwidth]{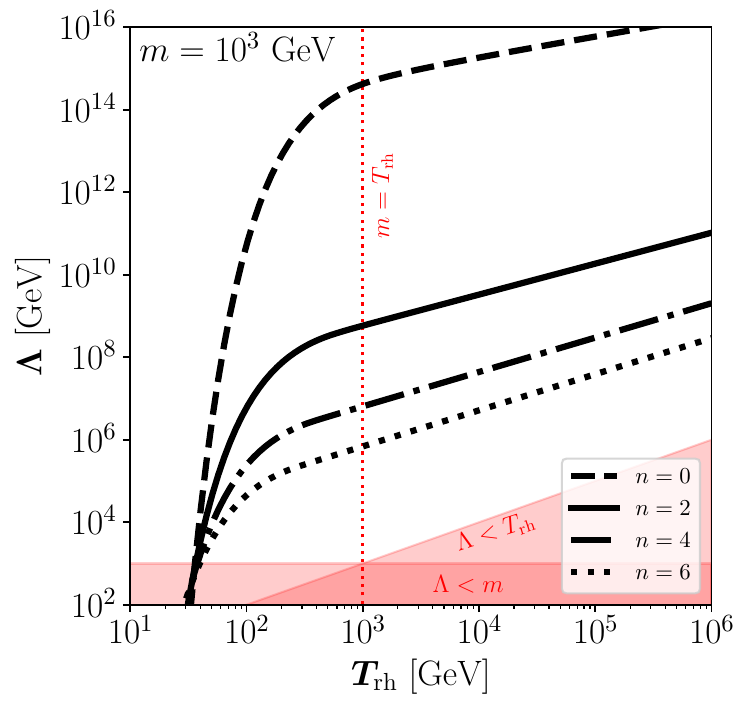}
    \caption{{\bf Instantaneous reheating.} As Figure~\ref{fig:plots} but for different values of $n$. With $n=0$, 2, 4, 6 corresponding to freeze-in operators of dimension 5, 6, 7, 8, respectively. We check that along each curve the dark matter does not enter equilibrium with the visible sector. 
    \label{fig:scan}}
\end{figure} 
%%%%%%%%%%%%%%%%%%%%%%%%%%%%%%%%%%%

The dark matter relic abundance due to the given UV freeze-in operator assuming instantaneous reheating is controlled by four parameters: $\Lambda$, $m$, $\Trh$, and $n$. In Figure~\ref{fig:plots} we consider the $s$-wave case ($L=0$) and sequentially fix $n$ with one other parameter to show the relationship between the remaining two free parameters that reproduces the observed dark matter relic abundance. For Figure~\ref{fig:plots}, it is assumed a mass dimension six UV freeze-in operator, corresponding to $n = 2$. In Figure~\ref{fig:scan} we generalize this to $n = 0$, 2, 4, 6, i.e.~operators of mass dimensions 5, 6, 7, 8. 

We overlay Figures~\ref{fig:plots} \&~\ref{fig:scan} with shaded red regions in which the EFT is not well defined: $\Lambda< m,~\Trh$. For $\Lambda > m,~\Trh$ one must define the UV completion in order to reliably carry out calculations, and one is no longer in the UV freeze-in regime. We note in passing that the successful reproduction of Big Bang Nucleosynthesis (BBN) requires $\Trh \gtrsim 4$~MeV~\cite{Kawasaki:2000en, Hannestad:2004px, Cyburt:2015mya, deSalas:2015glj}. This restriction does not constrain our examples.

Additionally, for Figure~\ref{fig:plots} we also show the region in which the inter-sector coupling causes the dark matter to come into thermal equilibrium with the visible sector bath, labeled ``thermalization''. In this region, dark matter will subsequently undergo thermal freeze-out, and the freeze-in mechanism will not determine the relic abundance. Thus, this region is outside of our scenario of interest. Specifically, to avoid thermalization, we impose that at all temperatures the following condition holds\footnote{In a recent paper, two of the present authors (SM \& JU) applied an alternative thermalization constraint~\cite{Chanda:2025bpl}. In place of Eq.~\eqref{444}, it was required that $i)$ at all times $n_{\rm eq} \gg n_{\rm DM}$ and $ii)$ dark matter annihilation does not occur after dark matter freeze-in. We have confirmed that these conditions are consistent with the restriction from Eq.~\eqref{444} up to $\mathcal{O}(1)$ variations.}
\begin{equation}\label{444}
	\Gamma_{\rm SM \rightarrow DM} = n_{\rm eq}\, \langle \sigma v \rangle < H\,,
\end{equation}
where $\Gamma_{\rm SM \rightarrow DM}(T)$ corresponds to the dark matter production rate. Thus we require the dark matter production rate on the left-hand (LH) expressions to be smaller than the Hubble rate at all times, to avoid chemical equilibrium between the two sectors. We do not show the thermalization constraints in Figure~\ref{fig:scan} as they are different for each operator dimension. However, we confirm that the thermalization constraints do not overlap any of the curves.

Notably, from Figures~\ref{fig:plots} \&~\ref{fig:scan}  we can make our principal observation:
\begin{center}
	\indent \em For dark matter mass $m>\Trh$ the yield is not zero, rather it is exponentially suppressed allowing for the prospect of UV freeze-in with a low-scale cut-off scale $\Lambda$.
\end{center}

That UV freeze-in can be realized with a low-scale cut-off is interesting for multiple reasons. Firstly, the new physics that is integrated out above the cut-off could be collider accessible. Indeed, as can be seen in the figures, the new physics could potentially be TeV scale. Moreover, such low cut-offs can potentially lead to direct and indirect detection signals; these constraints can be calculated once the UV freeze-in operator is specified (and without the need to specify the UV completion). Thus, we observe that UV freeze-in at lower cut-off is the EFT counterpart of IR freeze-in ``at stronger coupling''~\cite{Cosme:2023xpa}.

%%%%%%%%%%%%%%%%%%%%%%%%%%%%%%%%%%%%%
\subsection{Non-instantaneous reheating}
%%%%%%%%%%%%%%%%%%%%%%%%%%%%%%%%%%%%%
In the preceding section, we observed that for $m>\Trh$ one is in the regime of Boltzmann suppressed UV freeze-in; however, this assumed the instantaneous reheating approximation. In this section, we shall see that the general picture still holds away from the instantaneous reheating approximation, with some mild deviation. We shall see that for $\Trh<m<\Tmax$ there is a milder suppression of $Y_0$, with a full exponential suppression that arises only for $\Tmax<m$. This assumes an early matter-dominated phase before reheating, and in Section~\ref{sec4} we discuss cases in which this behavior differs.

%%%%%%%%%%%%%%%%%%%%%%%%%%%%%%%%%%%%%%%%%%%%%%%%%%%
\begin{figure}[t!]
    \def\sepf{0.328}
    \centering
    \includegraphics[width=\sepf\columnwidth]{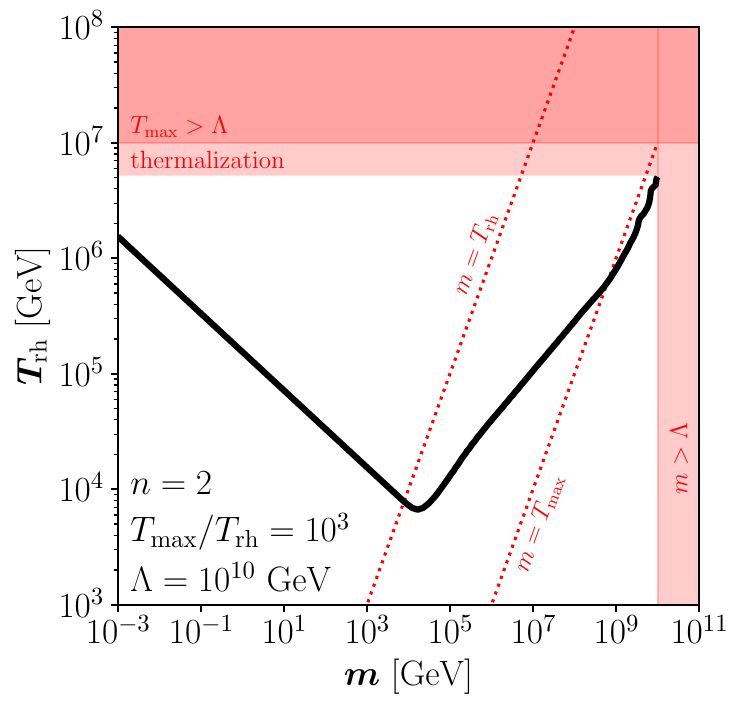}
    \includegraphics[width=\sepf\columnwidth]{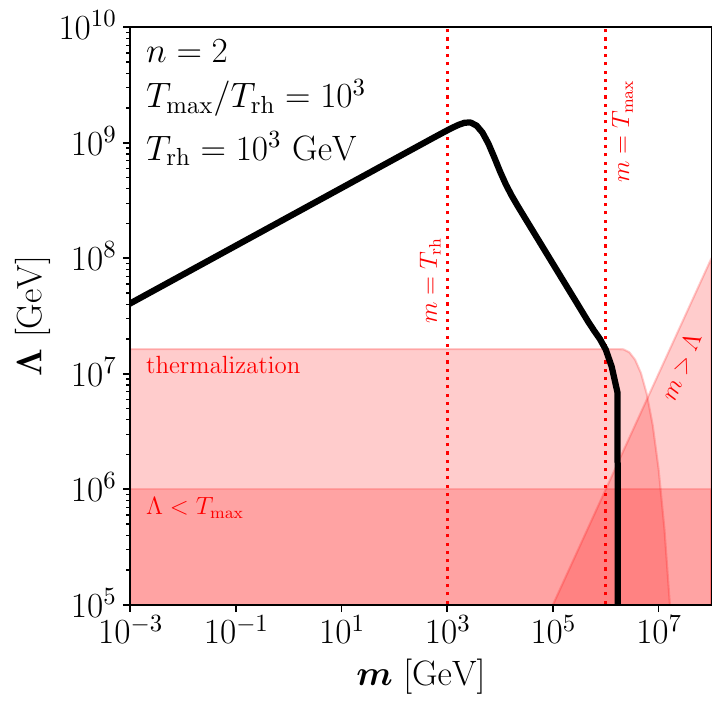}
    \includegraphics[width=\sepf\columnwidth]{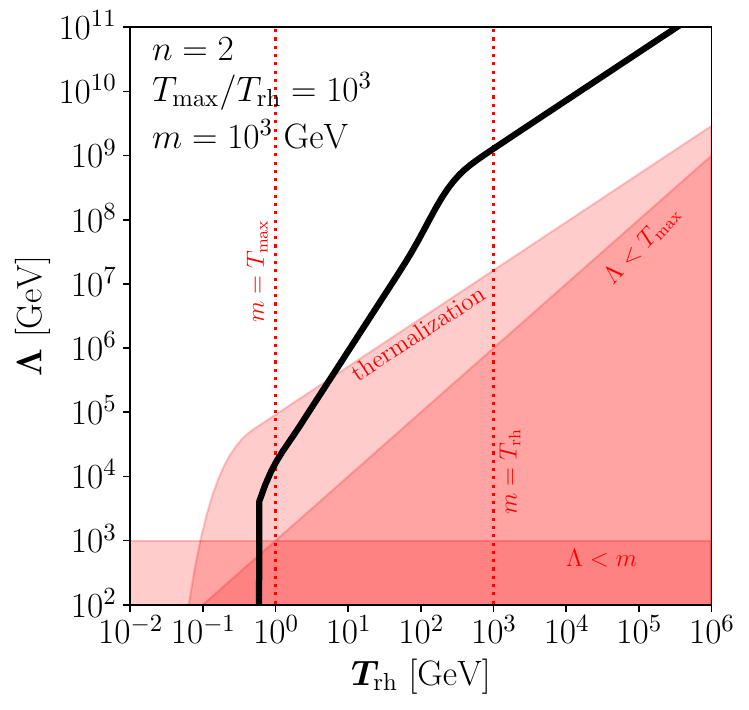}
    \caption{{\bf Non-instantaneous reheating.} The thick black line shows the parameter values for which the observed dark matter abundance is reproduced. The thin dashed correspond to the relativistic and non-relativistic approximations: black after reheating (that is, the same as in Figure~\ref{fig:plots}) and blue during reheating. The transition between the three regimes is depicted by the dotted red lines $m = \Trh$ and $m = \Tmax$. The red bands indicate $m > \Lambda$ or $\Tmax > \Lambda$. We assume $n=2$ and $L=0$, and that prior to reheating the Universe was in an early matter dominated phase.
    \label{fig:plots3}}
\end{figure}
%%%%%%%%%%%%%%%%%%%%%%%%%%%%%%%%%%%%%%%%%%%%%%%%%%%

If reheating is non-instantaneous, the highest temperature of the Universe $\Tmax$ can exceed $\Trh$~\cite{Giudice:2000ex}. In addition to dark matter production after reheating (leading to Eq.~\eqref{eq:Yafter}), production can also occur from interactions in the thermal bath during reheating. Moreover, in the reheating regime, Standard Model entropy is not conserved. Due to this, it is convenient to rewrite Eq.~\eqref{eq:BE0} as a function of $N(a) \equiv n_\text{DM}\, a^3$ and $a$
\beq \label{eq:Boltzmann}
    \frac{{\rm d}N}{{\rm d}a} = \frac{a^2}{H}\, n_\text{eq}^2\, \sv\,.
\eeq
It follows that freeze-in production of dark matter while inflaton decay is ongoing is given by the integral
\beq \label{eq:Yduring}
    Y_0 = \frac{45}{2\pi^2\, \gss(\Trh)}\, \frac{1}{\arh^3\, \Trh^3} \int_{a_*}^{\arh} \frac{a^2}{H(a)}\, n_\text{eq}^2\, \sv\, {\rm d}a\,,
\eeq
where $a=a_*$ corresponds to the scale factor at the beginning of reheating. For an operator of mass dimension $(n+2)$ and $n<6$ (we give the expression for general $n$ and $L=0$ in Section~\ref{sec4}, and $L>0$ in the Appendix), the yield is given by
\beq \label{eq:NIur}
    Y_0 \simeq \frac{180}{\pi^7\, \gss}\, \sqrt{\frac{10}{\gs}}\, \frac{\Mpl }{\Lambda^{2+n}} \times
	\begin{dcases}
    		\frac{1}{6-n}\,  \Trh^{1+n}  &\text{for } m\ll \Trh\,, \\
		      \frac{1}{2^{8}}\, \frac{\Trh^7}{m^{6-n}} \left[\Gamma\left(8, \frac{2\, m}{\Tmax}\right) - \Gamma\left(8, \frac{2\, m}{\Trh}\right)\right] &\text{for } m\gg \Tmax\,,
	\end{dcases}
\eeq
involving the incomplete Gamma function $\Gamma(l,x) = l!\, e^{-x} \sum_{k=0}^{l}\frac{x^{k}}{k!}$. The intermediate regime $\Tmax > m > \Trh$ can be estimated from the sum $Y_0|_{m\ll \Trh} + Y_0|_{m\gg \Trh}$ (from Eq.~\eqref{eq:NIur}), since both the relativistic and non-relativistic productions provide non-reliable contributions to $Y_0$.

%%%%%%%%%%%%%%%%%%%%%%%%%%%%%%%%%%%%%%%%%%%%%%%%
\begin{figure}[t!]
    \def\sepf{0.328}
    \centering
    \includegraphics[width=\sepf\columnwidth]{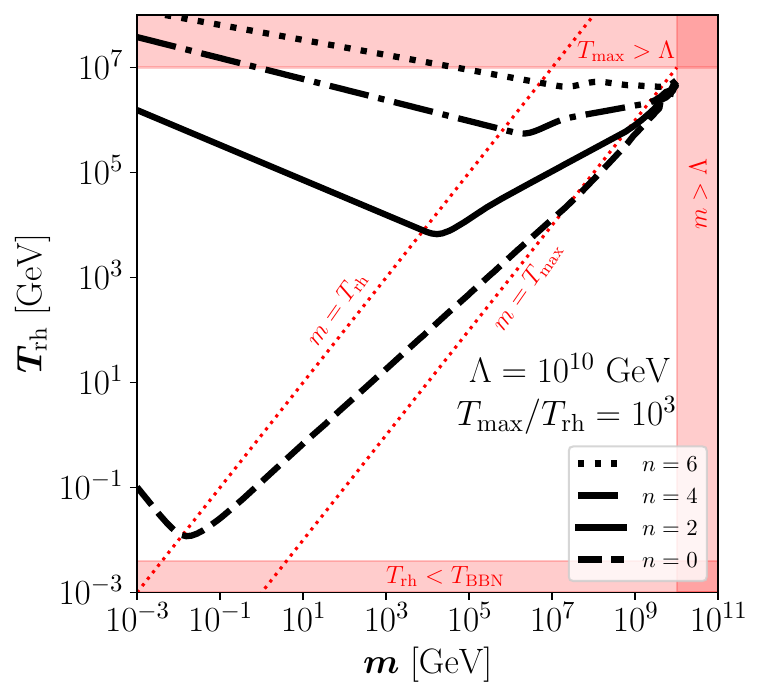}
    \includegraphics[width=\sepf\columnwidth]{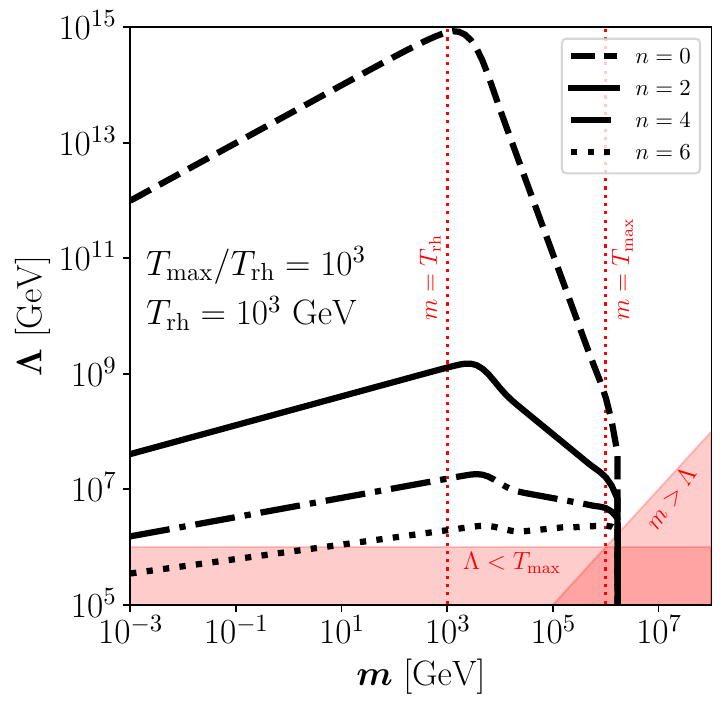}
    \includegraphics[width=\sepf\columnwidth]{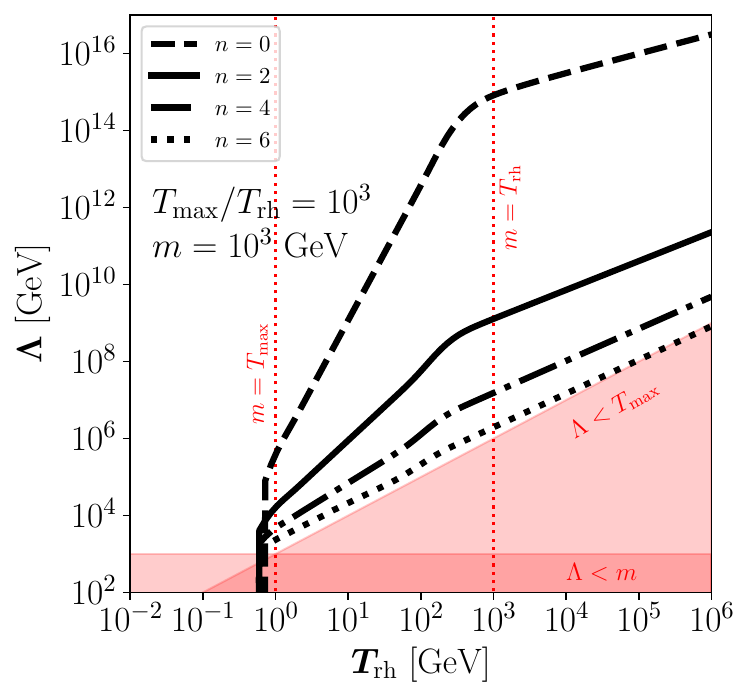}
    \caption{{\bf Non-instantaneous reheating.} As Figure~\ref{fig:plots3} but for different values of $n$ with $L=0$. With $n=0$, 2, 4, 6 corresponding to freeze-in operators of mass dimension 5, 6, 7, 8, respectively. We check that along each curve the dark matter does not enter equilibrium with the visible sector.
    \label{fig:plots4}}
\end{figure} 
%%%%%%%%%%%%%%%%%%%%%%%%%%%%%%%%%%%%%%%%%%%%%%%%%%%
In Figure~\ref{fig:plots3} we plot the parameter combinations that result in the correct dark matter abundance for the case $\Tmax=10^3\, \Trh$ with a dimension six operator ($n=2$) and assuming the freeze-in operator is generated at $\Lambda=10^{10}$GeV. Figure~\ref{fig:plots3} is the analog of Figure~\ref{fig:plots}, away from the instantaneous decay approximation. These parameter values are chosen to match Figure~\ref{fig:plots} (with the exception of $\Tmax$ which is not defined for instantaneous reheating). Similarly, Figure~\ref{fig:plots4}, is the non-instantaneous reheating analog of Figure~\ref{fig:scan}, and shows how $Y_0$ changes for operators of different mass dimension. We note that the $n=6$ curve is described by the more general equation, Eq.~\eqref{fff}, of Section~\ref{sec4}, since Eq.~\eqref{eq:NIur} is not valid for $n \geq 6$.

Comparing Figures~\ref{fig:plots} \&~\ref{fig:scan} and Figure~\ref{fig:plots3} \&~\ref{fig:plots4} we highlight that with $\Trh<\Tmax$ deviating away from the instantaneous reheating approximation results in modest changes in the dark matter abundance. In all cases the important takeaway is that for $n<6$ (i.e.~mass dimension $<8$), as the mass of dark matter is pushed above $\Tmax$, the scale required to reproduce the correct relic abundance drops sharply.  In the non-instantaneous case, suppression of the production rate also occurs for $m>\Trh$. This is best seen in the central panels of the figures.

For the variety of operator dimensions which we study and in the instantaneous and non-instantaneous case, it is the EFT restriction that bounds how low $\Lambda$ can be while realizing Boltzmann suppressed UV freeze-in. We highlight that for $\Trh=1$~TeV that 10~TeV dark matter with $\Lambda\sim100$~TeV seems attractive for linking UV freeze-in to the TeV scale and for generating potentially detectable signals.

%%%%%%%%%%%%%%%%%%%%%%%%%%%%%%%%%%%%%%%%%%%%%%%%%%%%%%%%%%
\section{Model Building} \label{sec3}
%%%%%%%%%%%%%%%%%%%%%%%%%%%%%%%%%%%%%%%%%%%%%%%%%%%%%%%%%%
One of the strengths of EFTs is that they can describe a multitude of UV completions. In this section we highlight a selection of well-motivated high-dimension operators which generically connect Standard Model states with dark matter. Furthermore, we present UV complete models (within the context of supersymmetry) and discuss how integrating out the heavier fields can lead to EFTs describing the interactions between the lightest supersymmetric particle and the Standard Model states. We relate each of these scenarios back to the model-independent analysis of Section~\ref{sec2}. We also highlight direct detection prospects where relevant.

%%%%%%%%%%%%%%%%%%%%%%%%%%%%%%%%%%%%%%%%%%%%%%
\subsection{The Higgs portal} \label{3.1}
%%%%%%%%%%%%%%%%%%%%%%%%%%%%%%%%%%%%%%%%%%%%%%
Cosme, Costa, and Lebedev's first study of Boltzmann suppressed IR freeze-in  took the renormalizable scalar dark matter Higgs portal as their prime example~\cite{Cosme:2023xpa}. Subsequent papers extended the scope of this study of the Higgs portal~\cite{Silva-Malpartida:2023yks, Arcadi:2024wwg, Lebedev:2024mbj, Khan:2025keb}. While the renormalizable Higgs portal is inconsistent with UV freeze-in,\footnote{For $m_h>\Tmax$ one can integrate out the Higgs leading to a dimension-six operator involving quarks and dark matter of the form $\frac{1}{\Lambda}\, \bar{q}\, q\, |\phi|^2$, with $\Lambda=m_h^2/(m_q\, \lambda)$. The issue is that $\Lambda$ controls both the dark matter relic density and the direct detection cross section. Reproducing the correct dark matter abundance while avoiding direct detection requires $\Trh\gtrsim m_h$ (and $m\gtrsim 3$~TeV)~\cite{Cosme:2023xpa} in which case the EFT is not valid.} the dimension-five Higgs portal for fermionic dark matter is an excellent candidate, as we explore below.

For the case of fermion dark matter $\psi$ through the Higgs portal, for $m_h<\Tmax$ there is no canonical UV completion, but there is the generic dimension-five operator~\cite{Biswas:2019iqm, Ikemoto:2022qxy, Haba:2024vdg, Arcadi:2024wwg, Lebedev:2024mbj, Mondal:2025awq}
\beq
    \frac{1}{\Lambda}\, |H|^2\, \bar{\psi}\, \psi\,.
    \label{1}
\eeq
Moreover, in cosmological calculations for $m_h>\Tmax$ (and in experimental/astrophysical calculations for $m_h>\sqrt{s}$) one can integrate out the Higgs leading to the dimension-six operator connecting quarks and fermion dark matter
\beq
    \frac{1}{\hat\Lambda^2}\, \bar{q}\, q\, \bar{\psi}\, \psi\,,
\eeq
where the EFT cutoff of the dimension-six operator $\hat\Lambda$ is related to the cutoff of Eq.~\eqref{1} via 
\beq
    \frac{1}{\hat\Lambda^2}=\frac{1}{\Lambda}\left(\frac{m_q}{m_h^2}\right).
\eeq  

These operators can be mapped to the model-independent analysis in Section~\ref{sec2}. Specifically, we note that the dimension-five and six operators correspond to $n=0$ and $n=2$ (in both cases these lead to $L=1$ final states).\footnote{The bilinear operators $|H|^2$, $\bar q q$ and $\bar\psi\psi$ are each Lorentz scalars with $J^{PC}=0^{++}$. Hence, both the dimension five and six operators have $J^{PC}=0^{++}$. For a fermion–antifermion pair $P=(-1)^{L+1}$, thus $L$ must be odd and the lowest allowed orbital angular momentum is $L=1$ ($p$-wave). Variant dimension five operators, such as $|H|^2\bar\psi\gamma^5\psi$, are CP even and thus $L=0$ ($s$-wave).} Characteristically, the parametric behavior closely follows the $n=0$ and $n=2$ curves in Figure~\ref{fig:scan} (neglecting the $L=1$ suppression leads to only small $\mathcal{O}(1)$ difference in the plots). The dimension six operator is also described by the curves in Figure~\ref{fig:plots}. As an example, let us consider the left panel of Figure~\ref{fig:plots}, this is indicative of an example in which dark matter is either a scalar (dashed line) or fermion (solid line) and the Higgs portal generated by new physics at the intermediate scale $10^{10}$~GeV (perhaps linked to neutrino masses or the PQ mechanism~\cite{Peccei:1977hh, Peccei:1977ur}).

%%%%%%%%%%%%%%%%%%%%%%%%%%%%%%%%%%%%%%%%%%%%%%%%%%%
\begin{figure}[t!]
    \def\sepf{0.45}
    \centering
    \includegraphics[width=\sepf\columnwidth]{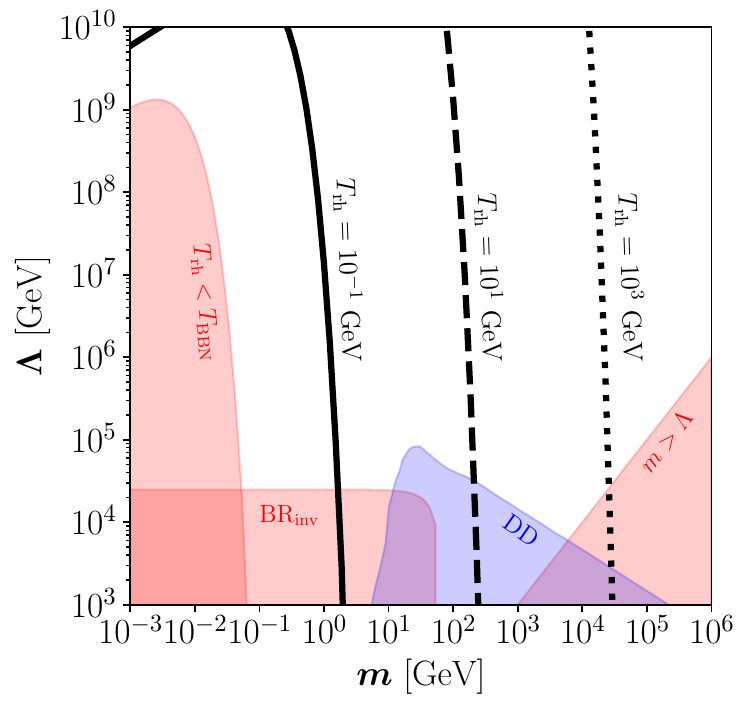}
    \includegraphics[width=\sepf\columnwidth]{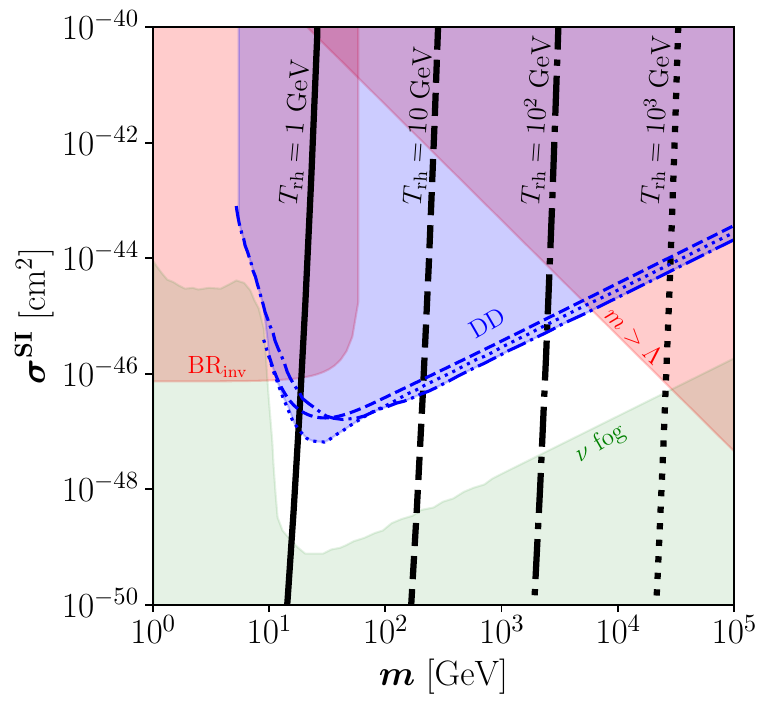}
    \caption{{\bf Dimension-five Higgs  portal.} Left. We consider fermion dark matter undergoing Boltzmann suppressed UV freeze-in via the Higgs portal, i.e.~$\frac{1}{\Lambda}|H|^2\bar{q}q$ leading to $\frac{m_q}{\Lambda m_h^2}\bar{\chi}\chi\bar{q}q$. Taking the instantaneous reheating approximation, and including the $p$-wave ($L=1$) suppression (cf.~Appendix~\ref{sec:app}), we show contours of $\Trh$ that give correct relic density of dark matter as $m$ and $\Lambda$ are varied. Experimental bounds from direct detection ``DD'' and invisible Higgs decay ``BR$_{\rm inv}$'' are overlaid. The shaded region indicates where EFT breaks down. We confirm that the dark matter does not equilibrate with the Standard Model over the relevant parameter space. Right.~We recast the LH panel in the familiar direct detection parameter space, showing the exclusion on the spin-independent scattering cross section from XENONnT~\cite{XENON:2025vwd}, PandaX-4T~\cite{PandaX:2024qfu}, and LUX-ZEPLIN~\cite{LZ:2022lsv}.  We also indicate the ``neutrino fog'' (green), where direct detection becomes challenging. 
    \label{fig:new}}
\end{figure} 
%%%%%%%%%%%%%%%%%%%%%%%%%%%%%%%%%%%%%%%%%%%%%%%%%%%
The spin-independent nucleon scattering cross section for fermion dark matter $\psi$ coupled to quarks via the Higgs portal $\frac{m_q}{\Lambda\, m_h^2}\, |\psi|^2\, \bar{q}\, q$ is given by~\cite{Kurylov:2003ra}
\begin{equation}
    \sigma_{\psi,{\rm SI}}^{(N)} = \frac{\mu^2}{\pi} \left(\frac{m_N}{\Lambda\, m_h^2}\, f_N^{(h)}\right)^2,
\end{equation}
where $\mu = \frac{m\,m_p}{m+m_p}$ is the dark matter-proton reduced mass (note that $\mu\simeq m_p$ for $m\gg m_p$) and $f_N^{(h)}\simeq0.3$ is the effective dark matter nucleon coupling. The spin-independent cross section of dark matter has been constrained by direct detection experiments, with the current leading bounds coming from XENONnT~\cite{XENON:2025vwd}, PandaX-4T~\cite{PandaX:2024qfu}, and LUX-ZEPLIN~\cite{LZ:2022lsv}. In Figure~\ref{fig:new} we combine these leading constraints into a single exclusion region, labeled ``DD''.

Furthermore, in the case where $m_h > 2\, m$ the Higgs can decay to dark matter states. Accordingly, collider constraints on the invisible Higgs width constrain the combination of $\Lambda$ and $m$. Specifically, the LHC bound on the invisible Higgs branching ratio is ${\rm BR}_{\rm inv} < B_{\rm inv} = 0.107$~\cite{ATLAS:2023tkt, CMS:2023sdw}, which can be mapped onto a constraint on the partial width
\beq
    \Gamma_{h\to\psi\bar\psi} < \Gamma_{\rm lim} \equiv \frac{B_{\rm inv}}{1 - B_{\rm inv}}\, \Gamma_h^{\rm tot} \simeq 4.9 \times 10^{-4}~{\rm GeV},
\eeq
where we have used the total Higgs width as $\Gamma_h^{\rm tot} \simeq 4.07$~MeV. For Dirac fermion dark matter $\psi$ of mass $m<m_h/2$ the partial Higgs width is
\beq
    \Gamma_{h\to\psi\bar\psi}=\frac{v^{2}}{\Lambda^{2}}\,\dfrac{m_h}{8\pi}\left(1-\dfrac{4m^{2}}{m_h^{2}}\right)^{3/2}.
\eeq
This implies the lower limit on the EFT scale for $m<m_h/2$
\begin{equation} \label{eq:LambdaInvH}
    \Lambda \gtrsim (2.49\times 10^{4}\,\mathrm{GeV})\, \left(1-\frac{4m^{2}}{m_h^{2}}\right)^{3/4}.
\end{equation}
We show this exclusion limit on Figure~\ref{fig:new}, labeled ``BR${}_{\rm inv}$''. We reiterate that for $m>m_h/2$ there is no constraint from Higgs decays. 

There is also the possibility that dark matter pair annihilation in high density astrophysical environments may lead to indirect detection signals. However, in the case of fermion dark matter all annihilation channels are velocity suppressed (i.e.~$\langle\sigma v\rangle\propto v^2$), cf. Ref.~\cite{Arcadi:2024ukq}. As a result, these limits are not competitive with direct detection, and therefore we do not show them in Figure~\ref{fig:new}.

In Figure~\ref{fig:new} we show how these bounds constrain the parameter space for the dimension five and six Higgs portals. There are two remarkable points to highlight here:
\begin{itemize}
    \item Unlike the traditional Higgs portal WIMP scenario, Boltzmann suppressed UV freeze-in via the Higgs portal is not experimentally excluded.
    \item In contrast to conventional UV freeze-in, the theory can be constrained (and potentially discovered) via direct detection.
\end{itemize}

%%%%%%%%%%%%%%%%%%%%%%%%%%%%%%%%%%%%%%%%%%%%%%%%%%%%%%%%%%%
\subsection{Supersymmetry above the reheating temperature} \label{3.3}
%%%%%%%%%%%%%%%%%%%%%%%%%%%%%%%%%%%%%%%%%%%%%%%%%%%%%%%%%%%
In supersymmetric models with R-parity conservation, particles have even R-parity, while sparticles have odd R-parity. Conservation of R-parity implies that sparticle decays must be to an odd number of superpartner final states; as a result, the lightest supersymmetric particle (LSP) is stable. Moreover, superpartners must be created in pairs, which has implications for the form of the freeze-in process. In the case $m_{\rm LSP}>\Tmax$, this state will undergo a Boltzmann-suppressed UV freeze-in with the higher-dimensional operators arising from integrating out the heavier superpartners. Here we consider the cases in which the LSP is a mostly bino neutralino, motivated by the CMSSM.

Despite a lack of positive evidence at the LHC, supersymmetry remains the leading candidate theory for resolving the hierarchy problem~\cite{ATLAS:2024lda}. It is also notable that in the Minimally Supersymmetric Standard Model (MSSM) with no large loop corrections, the Higgs mass is required by supersymmetry to be $m_h\simeq m_Z$. The Higgs mass can be increased to the observed value by including sizable loop corrections, and the simplest manner to achieve this is multi-TeV scale superpartners. While this introduces some mild fine-tuning in the Higgs sector, it opens an intriguing prospect that the superpartners may be heavier than the highest temperature of the thermal bath. In such a scenario, the LSP is an ideal dark matter candidate if it is electrically neutral and satisfies other direct searches~\cite{ATLAS:2024lda}. There are numerous candidates for dark matter in supersymmetric models; thus, let us consider a special and motivated case: the Constrained MSSM (CMSSM).

The CMSSM reduces the (large) number of new free parameters of the MSSM to just four continuous parameters and the sign of the Higgs sector $\mu$ term. Specifically, the CMSSM assumes that at the GUT scale the model has a single universal soft-breaking mass for the scalars $m_0$, a single universal gaugino mass $m_{1/2}$, and universal supersymmetry-breaking trilinear A-terms $A_0$. Thus, the only other deviations between sparticle masses and couplings arise from running from the GUT scale. The fourth free parameter is $\tan\beta$ being the ratio of the two Higgs VEVs. In particular, these conditions are not {\em ad hoc} simplifications, but consequences of generic gravity-mediated supersymmetry breaking.

In the CMSSM the LSP is typically a mostly bino neutralino $\chi$ or a stau $\tilde{\tau}$ (the gravitino may also be the LSP, as we consider in Section~\ref{3.4}).\footnote{\label{fn7} Deviating away from the CMSSM structure, the LSP could be a variety of different states, some of which are viable dark matter candidates. Two possibilities of note are the mostly Higgsino neutralino and the ``well-tempered'' neutralino~\cite{Arkani-Hamed:2006wnf}. Each of these variant dark matter candidates will have different implications for freeze-in and detection prospect, and would require a dedicated paper.} Which of the bino or stau is lightest (and thus the LSP) is determined by the relative sizes of $m_{0}$ and $m_{1/2}$, with the masses being given by~\cite{Martin:1993ft}
\beq
    m_\chi \approx 0.4\, m_{1/2} \qquad \text{and} \qquad m_{\tilde{\tau}}^2 \approx m_0^2 + 0.15\, m_{1/2}^2\,.
    \label{mass}
    \eeq
The stau carries electric charge and thus is not a viable dark matter candidate. Thus, we will require $m_0 > m_{1/2}/10$ such that the lightest neutralino is the LSP.  

%%%%%%%%%%%%%%%%%%%%%%%%%%%%%%%%%%%%%%%%%%%%%%%
\begin{figure}[t!]
    \centering
    \includegraphics[width=0.55\columnwidth]{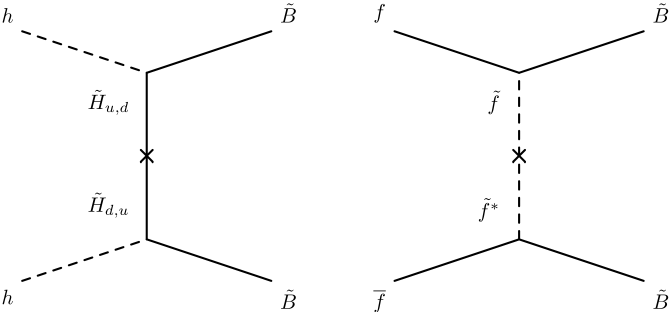}
    \caption{Relevant diagrams for bino freeze-in: Higgsino exchange (left) and sfermion exchange (right). Integrating out mediators leads to dimension five and six effective operators, respectively.}
    \label{Fig6}
\end{figure}
%%%%%%%%%%%%%%%%%%%%%%%%%%%%%%%%%%%%%%%%%%%%%%%
For simplicity, let us assume that the LSP can effectively be identified with a pure bino state. In this case, UV freeze-in process is typically set by $h + h \rightarrow \tilde{B} + \tilde{B}$ (see Figure~\ref{Fig6}, left) via a dimension five operator induced by integrating out heavy Higgsinos in the $t$-channel~\cite{Han:2023fgn}\footnote{Note that there is also a dimension-six operator $f + \bar{f} \rightarrow \tilde{B} + \tilde{B}$ (see Figure~\ref{Fig6}, right) from integrating out the heavy sfermions in the $t$-channel~\cite{Han:2023fgn}. This is suppressed by a factor $\propto m_{\tilde{f}}^2$ and therefore will generically be highly subleading unless $m_{\tilde{f}}\ll m_{\tilde{H}}$, which is not typical. For $\Tmax<m_h$, alternative dimension-6 operators that involve virtual Higgs loops can contribute, although these are suppressed by a loop factor $1/(16 \pi^2\, m_h^2)$.}
\beq
    \mathcal{L} \supset  &
    - \frac{g_1}{\sqrt{2}} (H_u^+)^* \tilde{H}_u^+ \tilde{B}
    - \frac{g_1}{\sqrt{2}} (H_u^0)^* \tilde{H}_u^0 \tilde{B}\\
&    + \frac{g_1}{\sqrt{2}} (H_d^-)^* \tilde{H}_d^- \tilde{B}
    + \frac{g_1}{\sqrt{2}} (H_d^0)^* \tilde{H}_d^0 \tilde{B}
    + \text{H.c.}
\eeq
At temperatures below electroweak symmetry breaking and after integrating out the Higgsinos one has
\beq
    \mathcal{L}_\text{eff} = \frac{g_1^2}{2\, \mu} \sin\beta\, \cos\beta\, h^2\, \tilde{B}\, \tilde{B}+ \text{H.c.}  
\eeq
Thus, the effective operator is dimension five with a cut-off scale
\beq
    \frac{1}{\Lambda} = \frac{g_1^2}{2\, \mu}\, \sin\beta\, \cos\beta\,.
    \label{eq1}
\eeq
From the inspection of Eq.~\eqref{eq1} we see that the effective operator formed by integrating out the Higgsinos is suppressed by the scale $\mu$, which occurs in the term $\mu\, H_u\, H_d$ and is anticipated to be of order $\mu \sim \tilde{m}$. Note that even for modest $3 \lesssim \tan\beta \lesssim 10$, one has $\sin\beta\, \cos\beta \sim 0.1$, which implies $\Lambda \sim 0.01\, \mu$. As we have seen in previous examples, this form of dimension five operator corresponds to the $n=0$ (and $L=1$) case; thus we can compare to the dashed curve in Figure~\ref{fig:scan}.

We note that in contrast to the Higgs portals, we anticipate that (being a gauge singlet) the bino would remain challenging to detect via direct or indirect detection searches, and for brevity we omit the exclusion limits (for relevant discussion, see Ref.~\cite{Han:2023fgn}). However, this conclusion does not extend to other LSP candidates, with Higgsino being an interesting alternative (see footnote~\ref{fn7}) which could lead to better detection prospects.

%%%%%%%%%%%%%%%%%%%%%%%%%%%%%%%%%%%%%%%%%%%%%%%%%%%%%%%%%%%%%%%%%%%%%%%%%%
\subsection{Gravitino dark matter} \label{3.4}
%%%%%%%%%%%%%%%%%%%%%%%%%%%%%%%%%%%%%%%%%%%%%%%%%%%%%%%%%%%%%%%%%%%%%%%%%%
A natural choice for UV freeze-in models is to fix the cut-off scale at the Planck scale $\Lambda = \Mpl$. Not only is this the only scale above the electroweak scale that is definitively identified in known physics, namely Newton's constant, but it is moreover generally expected that gravitational interactions should connect otherwise separated sectors of a theory. In particular, suppose that two sectors are decoupled due to the imposition of global symmetries; then gravitational interactions are expected to violate these global charges and allow the formation of any gauge-invariant operator. Due to these arguments, gravity-mediated UV freeze-in is particularly attractive. In fact, we highlight the particularly relevant work on gravity-mediated UV freeze-in at low reheating temperature; e.g. Refs.~\cite{Bernal:2018qlk, Lee:2024wes}. 

Aside from a separate sector, one particularly well motivated dark matter candidate that interacts gravitationally is the superpartner of the graviton, the spin-$3/2$ gravitino. In this section, we discuss the implications for gravitino production in the Boltzmann suppressed regime (the gravitino case was briefly discussed in Ref.~\cite{Giudice:2000ex}).

In the case where supersymmetry breaking is mediated to the visible sector via gravity, the mass scale of the other superpartners $\tilde{m}$ is comparable $\tilde{m}\sim m_{3/2}$. As noted in the previous section, in the CMSSM (which naturally arises in gravity-mediated supersymmetry breaking), the LSP is typically either the bino or the stau~\cite{Martin:1993ft}, see Eq.~\eqref{mass}. However, certain UV constructions can also lead to the gravitino being the LSP. Specifically, in gravity-mediated supersymmetry breaking, the gravitino can be the LSP provided that the gauge kinetic function and the Kähler potential satisfy a certain requirement~\cite{Kersten:2009qk}. Moreover, in gauge-mediated scenarios in which supersymmetry breaking is communicated to the Standard Model superpartners at a scale $\Lambda \ll \Mpl$, the gravitino is generically the LSP since $\tilde{m} = F_X/\Lambda \gg m_{3/2}$, where $F_X$ is the non-vanishing F-term which breaks supersymmetry. In fact, gauge mediation naturally leads to the hierarchy of Eq.~\eqref{hh}.

We consider a similar set-up to Section \ref{3.3}, now with a gravitino LSP, with the hierarchy
\beq
    \Trh < m_{3/2} < \tilde{m}\,.
    \label{hh}
\eeq
The gravitino $\psi_\mu$ has known interactions set by the structure of supergravity
\beq
    {\cal L} \supset - \frac{1}{\Mpl} \left[\frac{i}{2}\, \psi_\mu\, \sigma^{\nu\rho}\, \sigma^\mu\, \bar\lambda^a\, F^a_{\nu\rho} + \frac{i}{\sqrt{2}}\, (D_\mu \phi_*)^\dagger\, \psi_\nu\, \sigma^{\mu}\, \bar\sigma^\nu\, \psi_* + {\rm H.c.} \right].
\eeq
The leading production mechanism will be pair production of goldstinos, the spin 1/2 longitudinal component of the gravitino. The goldstino is the degree of freedom that is eaten by the gravitino when supersymmetry is spontaneously broken.

The relevant goldstino interactions appear after substitution $\psi_\mu \to \sqrt{\frac23}\, \frac{1}{m_{3/2}}\, (\partial_\mu\chi)$. Moreover, after integrating out the gaugino states $\bar{\lambda}$ (since we assume $\Trh \ll m_\lambda^a \sim \tilde{m}$) one arrives at the effective Lagrangian\footnote{We assume a single source for supersymmetry breaking via a non-vanishing hidden sector F-term. More complicated supersymmetry breaking~(cf. Refs.~\cite{Cheung:2010mc, Craig:2010yf})  can lead to interesting variations, as in Ref.~\cite{Monteux:2015qqa}.}
\beq
    \mathcal{L}\supset\frac{1}{F_X^2}\, \bar{\psi}\, \sigma^{\mu\nu}\, \partial^\rho \psi\, F_{\mu\rho}\,F_{\nu}^{\rho},
    \label{dim8}
\eeq
leading to a mass for the gravitino  $m_{3/2} = F_X/(\sqrt{3}\, \Mpl)$. Note that the suppression for the goldstino is $F_X^2\simeq (m_{3/2}\, \Mpl)^2$ unlike the $\Mpl^4$ suppression that occurs for the spin-3/2 component of the gravitino. This operator allows for the pair production process $gg\rightarrow \psi\psi$. It follows that one can identify with a mass dimension eight operator of Eq.~\eqref{dim8} with cut-off
\beq
    \Lambda \sim \sqrt{F} \sim \sqrt{\sqrt{3}\, m_{2/3}\, \Mpl}\,.
\eeq

The cross section to pair produce relativistic gravitinos is parametrically~\cite{Benakli:2017whb, Dudas:2017rpa, Garcia:2017tuj}
\beq\label{f4}
    \langle \sigma v\rangle \sim \frac{2000}{\pi}\, \frac{T^6}{F_X^4} \sim \frac{2000}{3\pi}\, \frac{T^6}{m_{2/3}^4\, \Mpl^4}\,.
\eeq
This is consistent with the model-independent cross section of Eq.~\eqref{sv}, for $n=6$. In the case $\Trh\ll \tilde{m}$ only the Standard Model particles take part in gravitino production and as their superpartners are not part of the thermal bath. Gravitino pair production is the relevant cross section since the other sparticles are not kinematically accessible in this scenario, indeed with $\Trh<m_{3/2}<\tilde{m}$ even gravitino production will be Boltzmann suppressed.\footnote{Gravitino production is also $p$-wave for $\Trh<m_{3/2}$. This can be inferred from the fact that the operator has exactly one spatial derivative acting on the fermions. This implies a single power of the final relative momentum which maps onto $p_f/p_i=\beta_f$ and thus, $\sigma \propto\beta_f^2(p_f/p_i)=\beta_f^{3}$ which is identified as $L=1$ scaling.}

The relevant plots showing the parameter scaling of this scenario which yield a gravitino relic abundance consistent with the observed dark matter density (assuming instantaneous reheating) are shown in Figure~\ref{fig:scan} for $n=6$. We have seen in Eq.~\eqref{f4} that the gravitino pair-production cross section is $\propto 1/F^4$ and the same is true for the cross sections that dictate $2\rightarrow2$ scattering rates and pair-annihilation rates involving gravitinos and Standard Model states. Consequently, no constraints arise from direct or indirect detection experiments.

%%%%%%%%%%%%%%%%%%%%%%%%%%%%%%%%%%%%%%%%%%%%%%%%%%%%%%%%%%
\section{Non-Standard Cosmology and Kination} \label{sec4}
%%%%%%%%%%%%%%%%%%%%%%%%%%%%%%%%%%%%%%%%%%%%%%%%%%%%%%%%%%
Thus far we have focused on the standard cosmological framework that assumes that the inflaton energy density led to a matter-dominated Universe, with the energy subsequently transferred to the radiation bath, leading to a radiation-dominated Universe prior to BBN. It is further assumed that this radiation domination gave way to matter domination at temperatures around 1~eV (matter radiation equality), before eventually transitioning to dark-energy domination. We refer to this as ``standard cosmology''.

In this section we shall contemplate the possibility that the equation of state of the Universe prior to UV freeze-in (and radiation domination) was something different from matter domination. The implications of non-standard cosmologies for UV freeze-in have previously been studied in Refs.~\cite{Garcia:2017tuj, Bernal:2019mhf, Bernal:2025fdr} (see also Refs.~\cite{Allahverdi:2020bys, Batell:2024dsi} for a more general review of non-standard cosmology). Notably, deviating away from standard cosmology can have a significant impact on UV freeze-in, including in the case in which $m>\Trh$, as we shall see below. 

Suppose that during cosmic reheating, the equation-of-state (EoS) parameter of the Universe is $\omega$ and that the Standard Model temperature scales as a power-law. We call $\Trh$ and $\arh$ the Standard Model temperature and the scale factor that correspond to the onset of the radiation-dominated era. Therefore, the evolution of the Hubble expansion rate $H$ as a function of the scale factor $a$ is given by~\cite{Bernal:2024yhu, Bernal:2024jim, Bernal:2024ndy, Bernal:2025fdr, Banik:2025olw}
\beq
    H(a) \simeq \Hrh \times
    \begin{dcases}
        \left(\frac{\arh}{a}\right)^\frac{3(1+\omega)}{2} &\text{ for } a \geq \arh\,,\\
        \left(\frac{\arh}{a}\right)^2 &\text{ for } \arh \geq a\,,
    \end{dcases}
\eeq
while the Standard Model bath temperature evolves as
\beq
    T(a) \simeq \Trh \times
    \begin{dcases}
        \left(\frac{\arh}{a}\right)^\alpha &\text{ for } a \geq \arh\,,\\
        \left(\frac{\arh}{a}\right) &\text{ for } \arh \geq a\,.
    \end{dcases}
\eeq
Typically $\alpha\geq0$, however, in certain scenarios it is possible that $\alpha<0$ so that the temperature decreases at times preceding $\arh$~\cite{Co:2020xaf}. One can generalize the form of $\Tmax$ to accommodate the general scaling, by defining~\cite{Bernal:2024yhu}
\begin{equation} \label{star}
    T_* \simeq \Trh \left[\frac{90}{\pi^2\, \gs}\, \frac{H_*^2\, \Mpl^2}{\Trh^4}\right]^\frac{\alpha}{3 (1+\omega)}.
\end{equation}
The simplest case is $\alpha=0$ then $T_*=\Tmax=\Trh$, the other conventional possibility is $\alpha=3/8$ which recovers the normal form of $T_*=\Tmax$~\cite{Giudice:2000ex}. Note that with $\alpha<0$ then $T_* \leq \Trh$ and $T_*$ corresponds to the end of inflation (and the onset of the initial phase of the Universe with EoS $\omega$). Notably, defining $H_*\equiv H(T_*)$ the BICEP/Keck bound on the tensor-to-scalar ratio~\cite{BICEP:2021xfz} provides a constraint: $H_* \leq 2.0 \times 10^{-5}~\Mpl$.

For conventional inflation models, it is typically assumed that $\omega = 0$ and $\alpha = 3/8$~\cite{Giudice:2000ex}, since these correspond to matter domination, and thus we took this as the vanilla case in Section~\ref{sec2}. Variations in inflationary cosmology can lead to deviations. For example, inflation models based on $\alpha$-attractor~\cite{Kallosh:2013hoa, Kallosh:2013maa}, allow for fairly general EoS parameter values. In such models during reheating the inflaton oscillates in a monomial potential $V(\phi) \propto \phi^p$ with $p\geq 2$ leading to $\omega = (p-2)/(p+2)$~\cite{Turner:1983he}.  A notable example, which we explore further below, is the case of kinetic energy domination or ``kination''~\cite{Spokoiny:1993kt, Ferreira:1997hj}. Specifically, for $\omega > 1/3$, the energy density in the inflaton field dilutes faster than that in radiation. In this case, it is not necessary for the inflaton to be completely transferred to light degrees of freedom. The exponent $\alpha$ depends on how the inflaton energy is transferred to the radiation. For instance, in annihilation-driven reheating gives model-dependent exponents leading to $\alpha=9/(2(p+2))$ for $\phi^p$ potentials (without resonances)~\cite{Co:2020xaf, Garcia:2020wiy, Xu:2023lxw, Barman:2024mqo}. If the inflaton decays rapidly or the temperature is constant during reheating~\cite{Co:2020xaf, Barman:2022tzk, Chowdhury:2023jft, Cosme:2024ndc} then $\alpha =0$. 

%%%%%%%%%%%%%%%%%%%%%%%%%%%%%%%%%%%%%%%%%%%%%%%%%%%
\begin{figure}[t!]
    \def\sepf{0.328}
    \centering
    \includegraphics[width=\sepf\columnwidth]{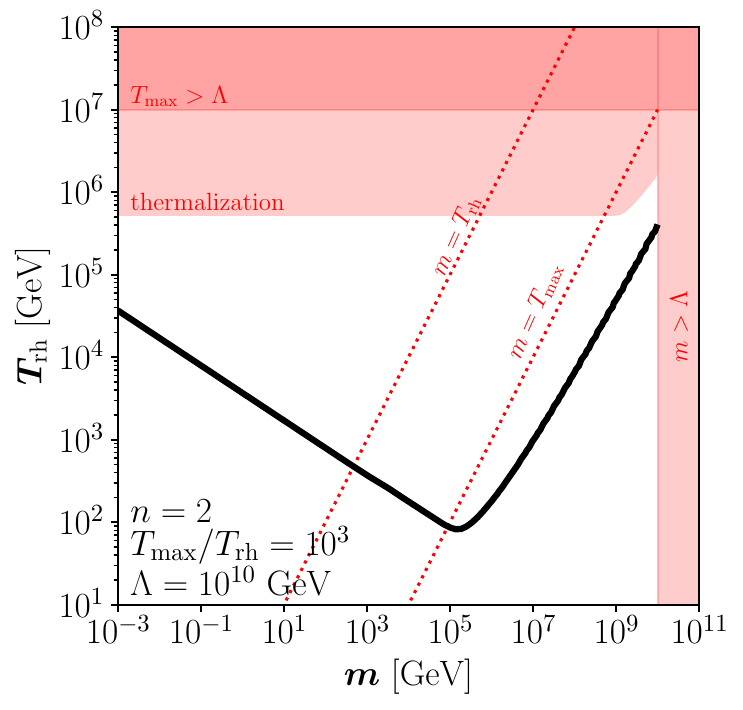}
    \includegraphics[width=\sepf\columnwidth]{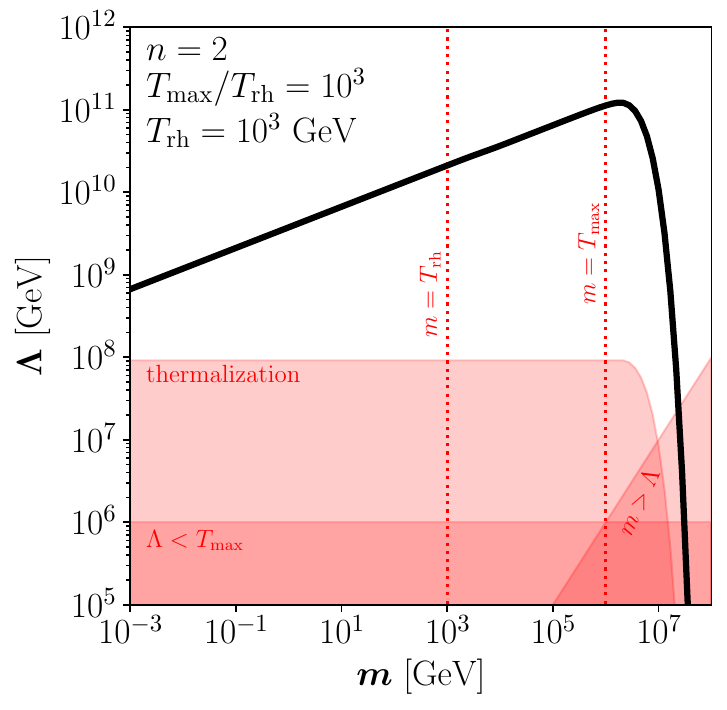}
    \includegraphics[width=\sepf\columnwidth]{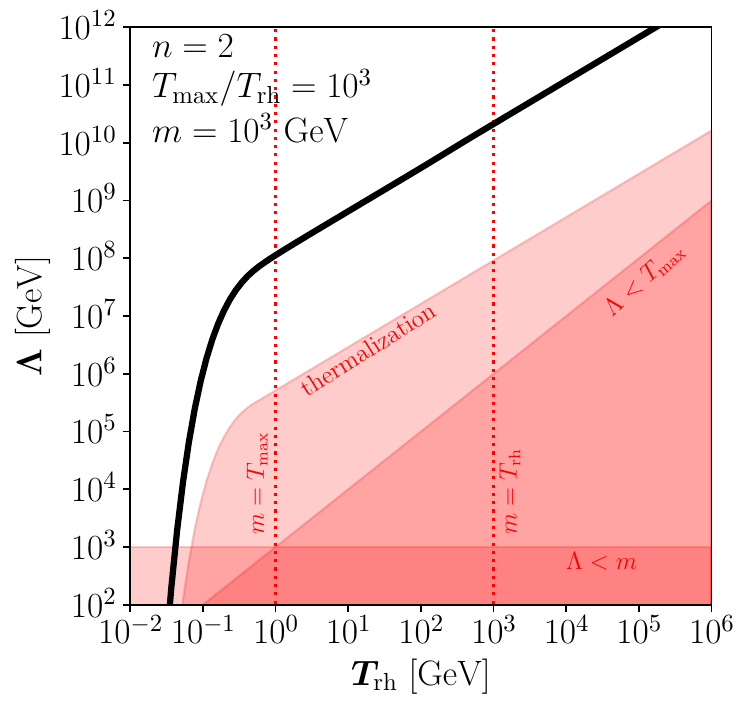}
    \caption{{\bf Kination: Non-instantaneous reheating.} Analogous to Figure~\ref{fig:plots3} (with $n=2$ and $L=0$) but for the case of kination domination  prior to reheating (i.e.~with $\omega = \alpha = 1$). \label{fig:plots5}}
    \end{figure}
%%%%%%%%%%%%%%%%%%%%%%%%%%%%%%%%%%%%%%%%%%%%%%%%%%%
    
As alluded to in Section~\ref{sec2}, the parametric forms of certain equations change as the dimension of the operators increases. In the case of matter domination (with $\omega=0$) the critical threshold is the mass dimension $n_c=6$. This critical threshold depends on $\omega$ and $\alpha$ and is given more generally by 
\beq\label{nc}
    n_c \equiv \frac32 \left( \frac{3+\omega-4\alpha}{\alpha}\right).
\eeq
We calculate the dark matter yield for operators above, below, and at this critical dimension, leading to three different parametric forms for the yield. Taking the relativistic limit $m \ll \Trh$ leads to
\beq\label{fff}
    Y_0 \simeq \frac{135}{2 \pi^7\, \gss}\, \sqrt{\frac{10}{\gs}}\, \frac{1}{\alpha}\, \frac{\Mpl\, \Trh^{1+n}}{\Lambda^{2+n}} \times
    \begin{dcases}
        \frac{1}{n_c - n} & \text{ for } n_c > n\,,\\
        \ln\left(\frac{T_*}{\Trh}\right) & \text{ for } n_c = n\,,\\
        \frac{1}{n - n_c} \left(\frac{T_*}{\Trh}\right)^{n-n_c} & \text{ for } n > n_c\,,
    \end{dcases}
\eeq
Whereas in the non-relativistic limit $m \gg T_*$ one has  (for $L=0$)
\beq\label{fff2}
    Y_0 &\simeq \frac{135}{2^{3+n_c}\, \pi^7\, \gss}\, \sqrt{\frac{10}{\gs}}\, \frac{1}{\alpha}\, \frac{\Mpl\, m^n\, \Trh}{\Lambda^{2+n}} \left(\frac{\Trh}{m}\right)^{n_c} 
 \left[\Gamma\left(2+n_c, \frac{2\, m}{T_*}\right) - \Gamma\left(2+n_c, \frac{2\, m}{\Trh}\right)\right].
\eeq
For early matter domination $n_c=6$ since $\omega=0$ and $\alpha = 3/8$; therefore, operators up to dimension 7 are described by the upmost cases of Eqs.~\eqref{fff} \& \eqref{fff2} (matching Eq.~\eqref{eq:NIur} in Section~\ref{sec2}).

Let us now look at the particular case of kination domination, in which the Universe is dominated by kinetic energy (for instance, of the inflaton field). This corresponds to $\omega=1$ and $\alpha=1$, according to $n_c=0$. Hence, in kination scenarios, $Y_0$ explicitly depends on $T_*$ for all UV freeze-in portals, regardless of the mass dimension of the portal operator. 

%%%%%%%%%%%%%%%%%%%%%%%%%%%%%%%%%%%%%%%%%%%%%%%%%%%
\begin{figure}[t!]
    \def\sepf{0.328}
    \centering
    \includegraphics[width=\sepf\columnwidth]{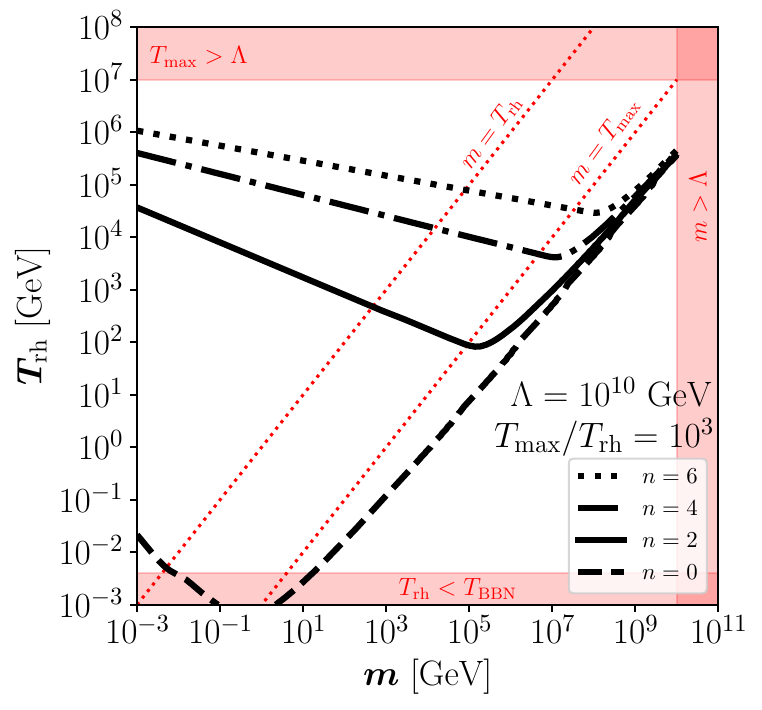}
    \includegraphics[width=\sepf\columnwidth]{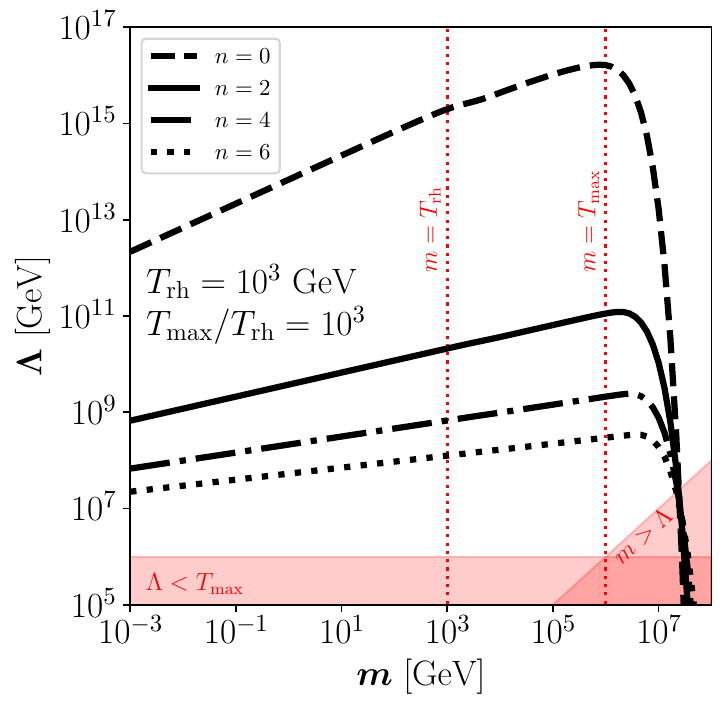}
    \includegraphics[width=\sepf\columnwidth]{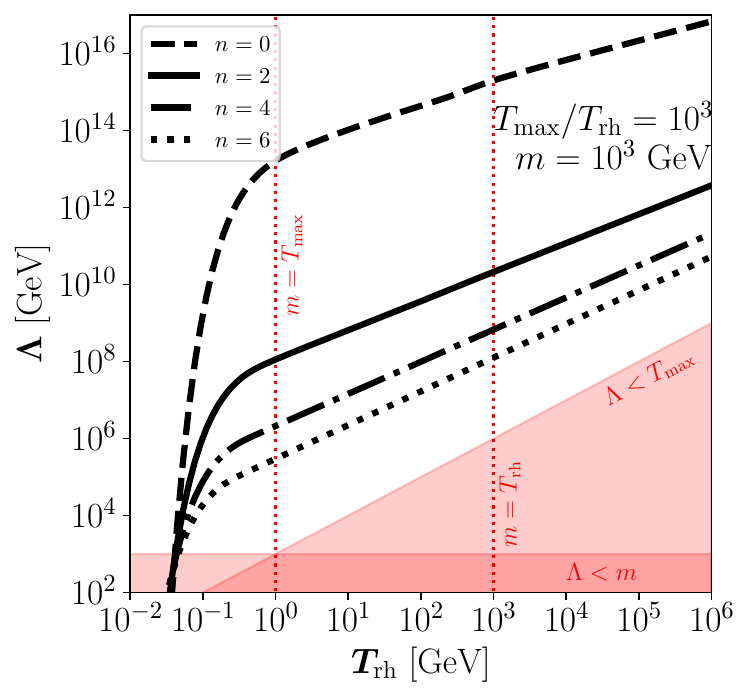}
    \caption{{\bf Kination: Non-instantaneous reheating.} Analogous to Figure~\ref{fig:plots4} (with $n=2$ and $L=0$) but for the case of kination domination prior to reheating (i.e.~with $\omega = \alpha = 1$). \label{fig:plots6}}
\end{figure} 
%%%%%%%%%%%%%%%%%%%%%%%%%%%%%%%%%%%%%%%%%%%%%%%%%%%
We specialize Eq.~\eqref{fff} \&~Eq.~\eqref{fff2} to the case of kination with ($n_c=0$) and consider $n\geq2$ (so mass dimension 6 and above) leading to
\beq\label{2222}
    Y_0 &\simeq \frac{135}{2 \pi^7\, \gss}\, \sqrt{\frac{10}{\gs}}\,   \frac{\Mpl\, \Trh}{\Lambda^{2+n}} 
  \times  \begin{dcases}
        \frac{1}{n}\,  T_*^n &\text{for } m\ll\Trh, \\
        \frac{1}{4}\, m^{n}  \left[\Gamma\left(2, \frac{2\, m}{\Tmax}\right) - \Gamma\left(2, \frac{2\, m}{\Trh}\right)\right]  &\text{for } m\gg \Tmax.
    \end{dcases}
    \eeq
Note, in particular, that the parametric dependence of the yield is significantly different from the matter-dominated analog of Eq.~\eqref{eq:NIur}. Figures~\ref{fig:plots5} \&~\ref{fig:plots6} provide the kination-domination analogs of the plots assuming early matter domination in Figures~\ref{fig:plots3} \&~\ref{fig:plots4}. Observe that, while the general structures are similar, there are some critical distinctions. In particular, we highlight that while for matter domination there is a strong suppression above $\Trh$, in the case of kination there is no suppression in the production until $m>\Tmax$, at which point the sharp exponential suppression arises; this is best seen in the central panel and is indicated by the necessary size of the cutoff $\Lambda$ needed to match observed relic density.

%%%%%%%%%%%%%%%%%%%%%%%%%%%%%%%%%%%%%%%%%%%%%%%%%%%%%%%%%%
\vspace{-2mm}
\section{Concluding Remarks} \label{sec5}
\vspace{-2mm}
%%%%%%%%%%%%%%%%%%%%%%%%%%%%%%%%%%%%%%%%%%%%%%%%%%%%%%%%%%
This work has presented the first dedicated analysis of Boltzmann-suppressed UV freeze-in. This is the EFT counterpart to the (IR) ``freeze-in at stronger coupling'' scenario~\cite{Cosme:2023xpa}. The important point is that for $m>\Trh$ the production rate is Boltzmann suppression, and one needs a radically smaller suppression scale in the freeze-in operator, compared to vanilla UV freeze-in (cf. Ref.~\cite{Elahi:2014fsa}). We also highlight that for $m>T$ the orbital angular momentum of the outgoing states $L$ can lead to an additional suppression factor. While the $L$-suppression tends to be mild compared to the exponential suppression, it is important for precision studies.

We have explored Boltzmann suppressed UV freeze-in for a range of operator dimensions (in Section~\ref{sec2}), in the context of different beyond-the-Standard Model scenarios (in Section~\ref{sec3}), and for varied cosmological assumptions (in Section~\ref{sec4}). In UV freeze-in for a given $\Trh$ the scale $\Lambda$ should be chosen so that the dark matter relic density is successfully reproduced. The key point of the Boltzmann suppressed UV freeze-in is that for $m>\max[\Trh,\Tmax]$ the necessary magnitude of the cut-off $\Lambda$ drops precipitately. This exponential suppression and the corresponding low-scale $\Lambda$ can be best observed in the central panels of Figs~\ref{fig:plots}-\ref{fig:plots4}, \ref{fig:plots5}, \& \ref{fig:plots6}.

In vanilla UV freeze-in to avoid thermalization between dark matter and the Standard Model bath, generically the cutoff scale in the portal operator $\Lambda$ is required to be high. Although connections to the PQ scale, the GUT scale, or the Planck scale are appealing, these scales are also experimentally inaccessible. Accordingly, the most exciting aspect of Boltzmann suppressed UV freeze-in is the fact that the scale of New Physics $\Lambda$ could be at the TeV scale making it directly accessible to next-generation collider experiments and potentially connected to the hierarchy problem (as in our supersymmetry scenario of Section~\ref{3.3}). Moreover, as we explored in Section~\ref{3.1} in the context of the Higgs portals, with low scale $\Lambda$ there is a prospect for direct detection signals of GeV-TeV scale freeze-in dark matter. 

Finally, we highlight that the Higgs portal has long been considered one of the best motivated manners of connecting dark matter to the Standard Model and it was a great disappointment to the community when this classic WIMP scenario was shown to be experimentally excluded. The prospect of reviving minimal and motivated portal operators within Boltzmann Suppressed freeze-in is an elegant prospect. Moreover, as can be seen in Figure~\ref{fig:new} (and for the renormalizable Higgs portal in Ref.~\cite{Cosme:2023xpa}), the framework of Boltzmann suppressed freeze-in presents a new opportunity to discover Higgs portal dark matter at direct detection experiments. In contrast, it also provides a new target for forthcoming direct detection experiments to strive to constrain in the near future.

%%%%%%%%%%%%%%%%%%%%%%%%%%%%%%%%%%%%%%%%%%%
\vspace{-2mm}
\acknowledgments
\vspace{-2mm}
%%%%%%%%%%%%%%%%%%%%%%%%%%%%%%%%%%%%%%%%%%%
NB received funding from the grants PID2023-151418NB-I00 funded by MCIU/AEI/10.13039 /501100011033/ FEDER and PID2022-139841NB-I00 of MICIU/AEI/10.13039/501100011033 and FEDER, UE. JU is supported by NSF grant PHY-2209998.

%%%%%%%%%%%%%%%%%%%%%%%%%%%%%%%%%%%%%%%%%%%
\appendix
%%%%%%%%%%%%%%%%%%%%%%%%%%%%%%%%%%%%%%%%%%%
%%%%%%%%%%%%%%%%%%%%%%%%%%%%%%%%%%%%%%%%%%%
\section{Dark Matter Production Cross Section} \label{sec:app}
%%%%%%%%%%%%%%%%%%%%%%%%%%%%%%%%%%%%%%%%%%%
In dark matter annihilation, one often takes a velocity expansion $\sigma v_{\rm rel} = a + b v_{\rm rel}^2 +\cdots$. In the case where $a=0$, this leads to $p$-wave suppression. This can be understood as arising from conservation of the orbital angular momentum of the incoming two-body system. Taking a partial-wave decomposition $a$ is the contribution of the $L=0$ partial wave, while $b v_{\rm rel}^2$ is the contribution of the $L=1$ ($p$-wave) partial wave. In standard freeze-in with high-reheating temperature, the Standard Model bath is relativistic, thus there is no such suppression since $v_{\rm rel}\simeq1$; however, in the case $T\ll m$ an analogous suppression can arise due to the orbital angular momentum of the outgoing two-body system. In this appendix, we derive the relevant expressions.

We consider the freeze-in process $B B\to\chi\bar\chi$ that involves massless bath particles $B$ and dark-matter mass particles of mass $m$ that interact via an operator of mass dimension $4+d$ (compared to the main text $d = 1+n/2$). For zero angular momentum in the final state, or when this effect can be neglected, one has the usual squared matrix element~\cite{Elahi:2014fsa}
\beq
    |\mathcal M(s)|^{2}\simeq\frac{s^{d}}{\Lambda^{2d}}\,.
\eeq
When the angular momentum of the final state cannot be neglected, as is often the case for $m\gg T$, this squared matrix element is modified as we discuss in the following.

Our starting point is the differential 2$\to$2 cross section (see, e.g.~PDG~\cite{ParticleDataGroup:2024cfk} \S49)
\beq
    \frac{{\rm d}\sigma}{{\rm d}\Omega} = \frac{1}{64\pi^2\, s}\, \frac{|\mathbf p_f|}{|\mathbf p_i|}\, |\mathcal M(s,\cos\theta)|^2.
\eeq
The production amplitude can be decomposed into orbital angular momentum modes  $L$
\beq
    \mathcal M(s,\cos\theta) = 16 \pi \sum_{L=0}^\infty (2L+1)\, a_L(s)\, P_L(\cos\theta)\,,
\eeq
where $P_L$ are the Legendre polynomials. The contribution of the leading allowed $L$ to the squared matrix element after integrating over the solid angle is~\cite{Wigner:1948zz, Quemener:2017bil}
\beq
    \big\langle|\mathcal M|^2\big\rangle_\Omega  \simeq \frac{s^{ d}}{\Lambda^{2d}}\, |a_L(s)|^2  \simeq \frac{s^d}{\Lambda^{2d}}\, \beta_f^{2L},
\eeq
with $\beta_f\equiv\sqrt{1-\frac{4m^2}{s}}$ and any $\mathcal{O}(1)$ factors absorbed into $\Lambda$. The occurrence of the suppression factor $\propto \beta_f^{ 2L}$ is a realization of {\em Wigner's threshold law}~\cite{Wigner:1948zz}. This is analogous to the calculation of two-body phase space near threshold (see, e.g., PDG~\cite{ParticleDataGroup:2024cfk} \S50).

It follows that for massless bath particles
\beq
    \sigma_L(s)=\frac{1}{16\pi\, s}\, \frac{|\mathbf p_f|}{|\mathbf p_i|}\, \big\langle|\mathcal M|^{2}\big\rangle_\Omega 
\eeq
with
\beq
    |\mathbf p_i|=\frac{\sqrt{s}}{2}\,,\qquad |\mathbf p_f|=\frac{\sqrt{s}}{2} \beta_f\,.
\eeq
Thus
\beq
    \sigma_L(s) = \frac{1}{16\pi\, s}\, \beta_f\, \Big[\frac{s^{ d}}{\Lambda^{2d}}\, \beta_f^{ 2L}\Big] = \frac{1}{16\pi}\, \frac{s^{ d-1}}{\Lambda^{2d}}\, \beta_f^{ 2L+1}\,.
\eeq
The thermally averaged production cross section can be expressed
\beq
    \langle\sigma v\rangle = \frac{\gamma_{ab\to\chi\bar\chi}}{n_{\rm eq}^2}
\eeq
in terms of the reaction density
\beq
    \gamma_{ab\to\chi\bar\chi}(T) = \frac{T}{32\pi^{4}} \int_{4m^2}^{\infty} {\rm d}s\, \sigma(s)\, (s-4m^{2})\, \sqrt{s}\, K_{1} \Big(\frac{\sqrt{s}}{T}\Big)\,,
\eeq
which we can evaluate in the non-relativistic and relativistic cases.

%%%%%%%%%%%%%%%%%%%%%%%%%%%%%%%%%%%%%%%%%%%%%%%%%%%%%%%%%%%%%%%%%%%
\subsection*{A.1\quad Relativistic $\boldsymbol{2\rightarrow 2}$  production}
%%%%%%%%%%%%%%%%%%%%%%%%%%%%%%%%%%%%%%%%%%%%%%%%%%%%%%%%%%%%%%%%%%%
We can take $m\approx 0$ in which case $\beta_f\approx 1$ (so no dependence on $L$) and
\beq
    \gamma(T) = \frac{ T}{32\pi^{4}}\, \frac{1}{16\pi\, \Lambda^{2d}} \int_{0}^{\infty} {\rm d}s\, s^{ d+\frac12}\, K_{1}\Big(\frac{\sqrt{s}}{T}\Big).
\eeq
With $x\equiv \sqrt{s}/T$ and ${\rm d}s = 2\, T^{2}\, x\, {\rm d}x$,
\beq
    \gamma(T) &= \frac{1}{256\pi^{5}\, \Lambda^{2d}}\, T^{2d+4} \int_{0}^{\infty} {\rm d}x\, x^{ 2d+2}\,K_{1}(x)\\
    &= \frac{1}{256\pi^{5}\, \Lambda^{2d}}\, T^{2d+4} \left[2^{ 2d+1}\, \Gamma(d+1)\, \Gamma(d+2)\right].
\eeq
Dividing by $n_{\rm eq}^2 \sim T^6/\pi^4$ yields 
\beq
    \langle\sigma v\rangle_{m\ll T} = \frac{2^{ 2d-7}}{\pi}\, d!\, (d+1)!\, \frac{T^{ 2d-2}}{\Lambda^{2d}}\,,
\eeq
which corresponds to the standard result~\cite{Elahi:2014fsa}. For instance, the dimension six case ($d=2$, thus $n=2$ in the dimension counting of the main text) is as expected
\beq
    \langle\sigma v\rangle_{m\ll T}\Big|_{n=4} = \frac{3}{2\pi}\, \frac{T^{2}}{\Lambda^{4}}\,.
\eeq

%%%%%%%%%%%%%%%%%%%%%%%%%%%%%%%%%%%%%%%%%%%%%%%%%%%%%%%%%%%%%%%%%%%
\subsection*{A.2\quad Non-relativistic $\boldsymbol{2\rightarrow 2}$ production}
%%%%%%%%%%%%%%%%%%%%%%%%%%%%%%%%%%%%%%%%%%%%%%%%%%%%%%%%%%%%%%%%%%%
Next we look at the non-relativistic limit, relevant for $T\ll m$, in this case we need the $\beta_f$ dependence in $\mathcal{M}$, and the reaction is
\beq \label{eq:new}
    \gamma(T)=\frac{ T}{32\pi^{4}} \int_{4m^2}^{\infty} {\rm d}s\, \sigma_L(s)\, (s-4m^2)\, \sqrt{s}\, K_{1} \Big(\frac{\sqrt{s}}{T}\Big).
\eeq
We restrict ourselves to the case of $2\rightarrow 2$ processes.\footnote{Generalizing the cross section to $2\rightarrow N$ processes is more complicated than a simple numerical prefactor. Rather, taking $N>2$ alters the parametric suppression of the production rate. The kinematics is complicated, and thus we shall leave this for future work.} To proceed we define the difference $\Delta\equiv \sqrt{s}-2m$ and write $\sqrt{s}=2m+\Delta$ with $\Delta\ge 0$ but small. Then
\beq
    s = 4\, m^2\, (1+\varepsilon)\,,
\eeq
with $\varepsilon = \frac{\Delta}{m}\left(1+\frac{\Delta}{4m}\right)\ll1$. It follows that
\beq
    {\rm d}s = 4\, m^2\, {\rm d}\varepsilon\,, \qquad \beta_f = \sqrt{\frac{\varepsilon}{1+\varepsilon}} \simeq \sqrt{\varepsilon}\,.
\eeq
To leading order in $\varepsilon$ we have 
\beq
    s^{ d-1} \simeq (4\, m^2)^{d-1}, \quad (s - 4\, m^2) \simeq 4\, m^2\, \varepsilon, \quad \sqrt{s}\simeq 2\, m\, .
\eeq
Thus
\beq
    \sigma_L(s) = \frac{1}{16\pi}\, \frac{s^{ d-1}}{\Lambda^{2d}}\, \beta_f^{ 2L+1} = \frac{1}{16\pi\, \Lambda^{2d}}\, (4\, m^2)^{d-1}\, \varepsilon^{L+\frac12}\,.
\eeq
We use $K_{1}(z) \simeq \sqrt{\pi/(2\, z)}\, e^{-z}$ for  large  $z$, thus
\beq
    K_{1}\left(\frac{\sqrt{s}}{T}\right) \simeq \sqrt{\frac{\pi\, T}{4\, m}}\, e^{-\frac{2\, m}{T}}\, e^{-\frac{m}{T}\, \varepsilon}.
\eeq
Thus, Eq.~\eqref{eq:new} can be re-expressed as
\begin{align}
    \gamma(T) &\simeq \frac{ T}{32\pi^{4}} \int_{0}^{\infty} {\rm d}\varepsilon\, (4\, m^2) \left[\frac{1}{16 \pi\, \Lambda^{2d}}\, (4\, m^2)^{d-1}\, \varepsilon^{L+\frac12}\right]  (4\, m^2\, \varepsilon)\, (2\, m) \left[\sqrt{\frac{\pi\, T}{4\, m}}\,  e^{-\frac{2\,m}{T}}\, e^{-\frac{m}{T}\, \varepsilon} \right] \nonumber\\
    &= \frac{ T}{\pi^{4}}\, \frac{1}{16 \pi\, \Lambda^{2d}}\, m^{5}\, (4\, m^2)^{d-1} \sqrt{\frac{\pi\, T}{4\, m}}\, e^{-2m/T} \int_{0}^{\infty} {\rm d}\varepsilon\, \varepsilon^{L+\frac32}\, e^{-\frac{m}{T}\, \varepsilon}\,.
\end{align}
The integral can be recognized as the Laplace transform definition of the $\Gamma$-function, specifically $\int_{0}^{\infty} \varepsilon^{p}\, e^{-q\, \varepsilon}\, {\rm d}\varepsilon = \Gamma(p+1)\, q^{-(p+1)}$ with $ p =L + \frac32$, and $q=m/T$, and thus
\beq
    \gamma(T)= \frac{ T}{\pi^{4}}\, \frac{1}{16 \pi\, \Lambda^{2d}}\, m^{5}\, (4\, m^2)^{d-1}\, \sqrt{\frac{\pi\, T}{4\, m}}\, e^{-\frac{2m}{T}}\, \Gamma\Big(L+\frac{5}{2}\Big) \left(\frac{T}{m}\right)^{L+\frac52}.
\eeq
Then we can divide by $n_{\rm eq}^2=T^{6}/\pi^{4}$ to obtain the thermally averaged cross section
\beq\label{sva}
    \langle\sigma v\rangle = \frac{\gamma}{n_{\rm eq}^2} = \frac{4^{ d-1}}{32}\, \frac{1}{\sqrt{\pi}}\, \Gamma \Big(L+\frac{5}{2}\Big)\, \frac{m^{ 2d}}{\Lambda^{2d}\, T^2} \left(\frac{T}{m}\right)^{ L} e^{-\frac{2\, m}{T}}.
\eeq

Finally, we check the simplest case $L=0$, for which $\Gamma(5/2)=3\sqrt{\pi}/4$ and thus
\beq
    \langle\sigma v\rangle_{m\gg T}\Big|_{L=0} \simeq \frac{3}{32}\, 4^{d-2}\, \frac{m^{2d}}{\Lambda^{2d}\, T^{2}}\, e^{-\frac{2\, m}{T}}.
\eeq
The other case that commonly arises is $L=1$, for which $\Gamma(7/2)=15\sqrt{\pi}/8$ and
\beq
    \langle\sigma v\rangle_{m\gg T}\Big|_{L=1} = \frac{30}{32}\, 4^{ d-3}\, \frac{m^{ 2d}}{\Lambda^{2d}\, T^2} \left(\frac{T}{m}\right) e^{-\frac{2\, m}{T}}.
\eeq
Note the relative suppression of $\frac{5}{2}\left(\frac{T}{m}\right)$ between $L=1$ and $L=0$ processes. 

%%%%%%%%%%%%%%%%%%%%%%%%%%%%%%%%%%%%%%%%%%%%%%%%%%%%%%%%%%%%%
\subsection*{A.3 \quad Dark matter yield for $\boldsymbol{L>0}$}
\begin{figure}[t!]
   \def\sepf{0.328}
    \centering
    \includegraphics[width=8cm]{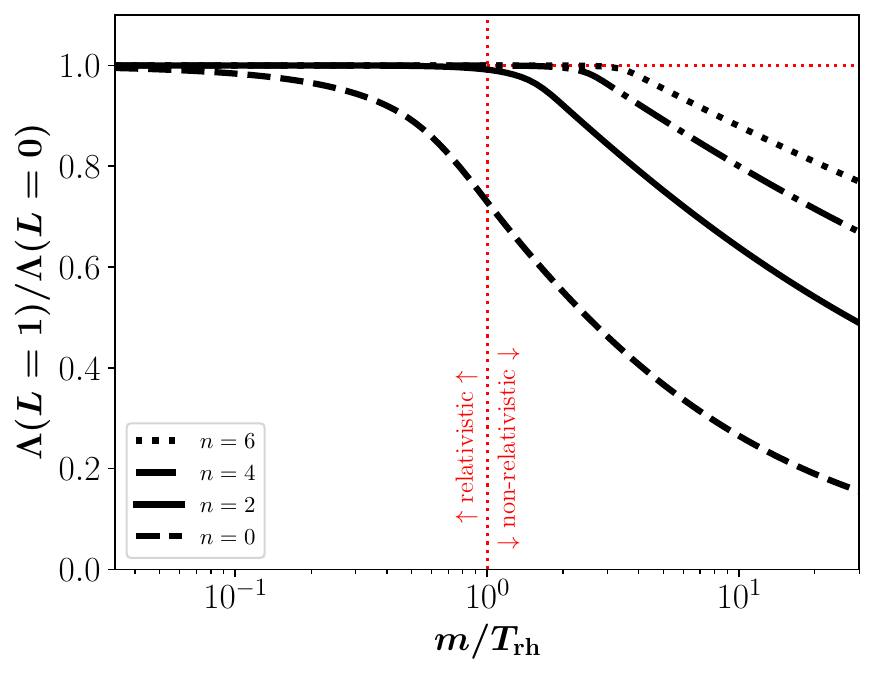}
\vspace{-2mm}    \caption{Suppression of the values of $\Lambda$ required to fit the entire observed DM abundance for $L=1$ with respect to the case where $L=0$.}
    \label{fig:reduction}
\end{figure}

%%%%%%%%%%%%%%%%%%%%%%%%%%%%%%%%%%%%%%%%%%%%%%%%%%%%%%%%%%%%%
Using the general forms of the production cross section derived above, we compute the dark matter yield. Starting with the case of instantaneous reheating for $L \geq 0$, we find
\beq \label{aa1}
    Y_0  =
    \begin{dcases}
        2^{n-6}\frac{135}{(n+1) \pi^8 \gss }\sqrt{\frac{10}{\gs}} \left(\frac{n+2}{2} \right)!  \left( \frac{n+2}{4} \right)! \, \frac{\Mpl\, \Trh^{n+1}}{\Lambda^{n+2}} & \text{for } m \ll \Trh\,,\\[3pt]
         \frac{135}{\pi^{15/2} \gss} \sqrt{\frac{10}{\gs}} 2^{n+L-7} \Gamma \left( L+\frac{5}{2} \right) \frac{\Mpl\, m^{n+1}}{\Lambda^{n+2}} \Gamma \left( 1-L, \frac{2\, m}{\Trh} \right) & \text{for } \Trh \ll m\,.
    \end{dcases}
\eeq
In the cross section stated in Eq.~\eqref{sv} we have absorbed the $\mathcal{O}(1)$ prefactors into $\Lambda$ compared to Eq.~\eqref{sva}. To obtain the form of Eq.~\eqref{eq:Yafter-app} one should divide the relativistic yield above by a factor of $c_{\rm rel} = 2^{n-5}/\pi \times ((n+2)/2)!((n+2)/4)!$ and in the non-relativistic case take $L=0$ and divide by $c_{\rm NR} = 2^{n-5}/\sqrt{\pi} \times \Gamma(5/2)$. The suppression of the dark matter yield due to higher values of $L$ in the incomplete Gamma function has to be compensated by an order-one reduction of the scale $\Lambda$. This effect can be observed in Figure~\ref{fig:reduction}, where the ratio of the values of $\Lambda$ required to fit the entire observed DM abundance for $L=1$ and $L=0$ is plotted for different values of $n$, as a function of $m/\Trh$. An order-one suppression appears only when DM is produced non-relativistically, and is maximized for low values of $n$.
%%%%%%%%%%%%%%%%%%%%%%%%%%%%%%%%%%%%%%%%%%%%%%%%%%%
%%%%%%%%%%%%%%%%%%%%%%%%%%%%%%%%%%%%%%%%%%%%%%%%%%%

Next we generalize the non-instantaneous reheating result of Eq.~\eqref{eq:NIur} assuming matter domination prior to reheating, that is $\omega = 0$ and $\alpha = 3/8$, to $L>0$ in which case we obtain 
\begin{align} \label{eq:NIurX}
    Y_0 &= \frac{180\times 2^{n-5}}{\gss}\, \sqrt{\frac{10}{\gs}} \nonumber\\
	&\times
    \begin{dcases}
    		\frac{1}{\pi^8 (6-n)} \left( \frac{n+2}{2} \right)! \left( \frac{n+4}{2} \right)! \frac{\Mpl\, \Trh^{n+1}}{\Lambda^{n+2}} \left[1 - \left(\frac{\Trh}{\Tmax}\right)^{6-n} \right] &\text{for } m\ll \Trh, \\
		      \frac{2^{-L-8}}{\pi^{15/2}} \frac{\Mpl\, \Trh^7\, m^{n-6}}{\Lambda^{n+2}} \left[\Gamma\left(8-L, \frac{2\, m}{\Tmax}\right) - \Gamma\left(8-L, \frac{2\, m}{\Trh}\right)\right] &\text{for } m\gg \Tmax.
	\end{dcases}
\end{align}
This reproduces Eq.~(\ref{eq:NIur}) if we restore the $\mathcal{O}(1)$ factors by dividing the relativistic yield by $c_{\rm rel}$, the non-relativistic yield by $c_{\rm NR}$, and take the limit $\Tmax \gg  \Trh$. 

Finally, we consider the generalization to the freeze-in yield non-standard cosmology in Eq.~\eqref{fff2} (with general $\omega$ and $\alpha$); in the non-relativistic case with $L>0$ we find
\begin{align}\label{fff2A}
    Y_0 &= \frac{135 \times 2^{n+L-n_c - 8}}{ \pi^{15/2}\, \gss}\, \sqrt{\frac{10}{\gs}}\, \frac{1}{\alpha}\, \frac{\Mpl\, m^n\, \Trh}{\Lambda^{2+n}} \left(\frac{\Trh}{m}\right)^{n_c} \nonumber\\
    &\qquad \times \Gamma\left( L + \frac{5}{2} \right) \times \left[\Gamma\left(2+n_c-L, \frac{2\, m}{T_*}\right) - \Gamma\left(2+n_c-L, \frac{2\, m}{\Trh}\right)\right].
\end{align}
Again, we recover the equation in the text by setting $L=0$ and dividing by $c_{\rm NR}$. 

%%%%%%%%%%%%%%%%%%%%%%%%%
\bibliographystyle{JHEP}
\bibliography{biblio}
%%%%%%%%%%%%%%%%%%%%%%%%%
\end{document}